\documentclass[dvips]{acta}
\usepackage{supertabular,lscape,epsfig}
\usepackage{amssymb}
\usepackage{amsmath}
\DeclareSymbolFont{ppa}{OT1}{ppl}{m}{it}
\DeclareMathSymbol{\vv}{\mathalpha}{ppa}{'166}

\newfont{\hb}{rphvb at 10pt}
\newfont{\hbo}{rphvbo at 10pt}
\newfont{\bitt}{rptmbi at 12pt}
\newfont{\bits}{rptmbi at 11pt}

\SetPages{67}{95}

\SetVol{62}{2012}

\begin{document}

\newcommand{\TabApp}[2]{\begin{center}\parbox[t]{#1}{\centerline{
  {\bf Appendix}}
  \vskip2mm
  \centerline{\small {\spaceskip 2pt plus 1pt minus 1pt T a b l e}
  \refstepcounter{table}\thetable}
  \vskip2mm
  \centerline{\footnotesize #2}}
  \vskip3mm
\end{center}}

\newcommand{\TabCapp}[2]{\begin{center}\parbox[t]{#1}{\centerline{
  \small {\spaceskip 2pt plus 1pt minus 1pt T a b l e}
  \refstepcounter{table}\thetable}
  \vskip2mm
  \centerline{\footnotesize #2}}
  \vskip3mm
\end{center}}

\newcommand{\TTabCap}[3]{\begin{center}\parbox[t]{#1}{\centerline{
  \small {\spaceskip 2pt plus 1pt minus 1pt T a b l e}
  \refstepcounter{table}\thetable}
  \vskip2mm
  \centerline{\footnotesize #2}
  \centerline{\footnotesize #3}}
  \vskip1mm
\end{center}}

\newcommand{\MakeTableApp}[4]{\begin{table}[p]\TabApp{#2}{#3}
  \begin{center} \TableFont \begin{tabular}{#1} #4 
  \end{tabular}\end{center}\end{table}}

\newcommand{\MakeTableSepp}[4]{\begin{table}[p]\TabCapp{#2}{#3}
  \begin{center} \TableFont \begin{tabular}{#1} #4 
  \end{tabular}\end{center}\end{table}}

\newcommand{\MakeTableee}[4]{\begin{table}[htb]\TabCapp{#2}{#3}
  \begin{center} \TableFont \begin{tabular}{#1} #4
  \end{tabular}\end{center}\end{table}}

\newcommand{\MakeTablee}[5]{\begin{table}[htb]\TTabCap{#2}{#3}{#4}
  \begin{center} \TableFont \begin{tabular}{#1} #5 
  \end{tabular}\end{center}\end{table}}

\newfont{\bb}{ptmbi8t at 12pt}
\newfont{\bbb}{cmbxti10}
\newfont{\bbbb}{cmbxti10 at 9pt}
\newcommand{\uprule}{\rule{0pt}{2.5ex}}
\newcommand{\douprule}{\rule[-2ex]{0pt}{4.5ex}}
\newcommand{\dorule}{\rule[-2ex]{0pt}{2ex}}
\newcommand{\meani}{$\langle I\rangle$}
\newcommand{\meanv}{$\langle V\rangle$}
\begin{Titlepage}
\Title{ASAS Photometry of ROSAT Sources. I.~Periodic Variable Stars 
Coincident with Bright Sources from the ROSAT All Sky Survey}
\Author{M.~~K~i~r~a~g~a}{Warsaw University Observatory, Al. Ujazdowskie 4,
00-478 Warszawa, Poland\\
e-mail: kiraga@astrouw.edu.pl}
\Received{February 28, 2012}
\end{Titlepage}

\Abstract{ 
Photometric data from the ASAS -- South (declination less than 29\arcd)
survey have been used for identification of bright stars located near the
sources from the ROSAT All Sky Survey Bright Source Catalog (RBSC). In
total 6028 stars brighter than 12.5~mag in {\it I}- or {\it V}-bands have
been selected and analyzed for periodicity. Altogether 2302 variable stars
have been found with periods ranging from 0.137~d to 193~d. Most of these
stars have X-ray emission of coronal origin with a few cataclysmic binaries
and early type stars with colliding winds. Whenever it was possible we
collected data available in the literature to verify periods and to
classify variable objects.

The catalog includes 1936 stars (1233 new) considered to be variable due to
presence of spots (rotationally variable), 127 detached eclipsing binary
stars (33 new), 124 contact binaries (11 new), 96 eclipsing stars with
deformed components (19 new), 13 ellipsoidal variables (4 new), 5
miscellaneous variables and one pulsating RR~Lyr type star (blended with an
eclipsing binary). More than 70\% of new variable stars have amplitudes
smaller than 0.1~mag, but for ASAS 063656-0521.0 we have found the largest
known amplitude of brightness variations due to the presence of spots (up
to $\Delta V=0.8$~mag). The table with the compiled data and figures with
light curves can be downloaded from the {\it Acta Astronomica
Archive}.}{Stars: variables: general -- Stars: rotation -- Stars: activity
-- X-rays: stars}

\vspace*{6pt}
\Section{Introduction}
The X-ray emission is related to high energy phenomena and its presence
usually points to astrophysically interesting objects. The stellar coronae
are the most ubiquitous X-ray sources in the solar neighborhood and their
existence is related to the magnetic field generation {\it via} dynamo
action below or within convective zones of cool stars (\eg Ossendrijver
2003, Brandenburg 2005, Browning 2008). The ROSAT all sky survey has
unveiled a wealth of information about stellar activity in the solar
surroundings (Voges \etal 1999). One of the most important parameters
governing magnetic field generation is stellar rotation, which usually
enters into studies of magnetic activity {\it via} the Rossby number: the
ratio of rotation period to the convective overturn time near the bottom of
convective zone (Noyes \etal 1984, Stêpieñ 1994, Pizzolato \etal 2003,
Wright \etal 2011). To fully exploit ROSAT data in the study of
activity \vs rotation period relation it is tempting to obtain as large as
possible number of rotational periods of the ROSAT X-ray sources. Their
photometric studies have been already subject of several papers (\eg
Szczygie³ \etal 2008, Norton \etal 2007), but complete survey of their
photometric variability has not been done yet.

The number of stars known to be variable due to rotation of the spotted
(nonuniform brightness) surface is growing quickly due to extensive
photometric studies of star clusters and star formation regions.

Recently, Hartman \etal (2011) published an extensive study of rotational
variability of K and M dwarfs based on the HATNet data for stars with
declination between $+15\arcd$ and $+52\arcd$. Selection of stars was based
on their colors and proper motions. The authors have found 2120 variable
stars out of 27\,560 stars analyzed for variability. Only 38 of these
variable stars had been known before.

Fully convective M dwarf stars were the subject to the recent study by
Irwin \etal (2011). The authors performed period search for 273 dwarfs and
found 41 variable M-dwarfs. It is interesting that they found several stars
with periods longer than 100~d, with the longest period equal to 154~d for
LHS1667.

Our study is based on photometric data from the All Sky Automatic Survey
(ASAS). A large catalog of about 50\,000 variable stars, available {\it
via} Internet has been described in several papers: Pojmañski 2002,
Pojmañski 2003, Pojmañski and Maciejewski 2004, Pojmañski and Maciejewski
2005, Pojmañski \etal 2005 (ASAS Catalog of Variable Stars: hereafter ACVS)
but the data obtained by ASAS have not been fully explored yet. It is
possible to select particularly interesting object or class of objects for
subsequent study (for example study of $\beta$~Cep variables: Pigulski
2005, or X-ray detected eclipsing binaries: Szczygie³ \etal 2008), and also
to find more variable stars not yet identified in the ASAS photometric
database, as it has been shown in the search for $\beta$~Cep stars
(Pigulski and Pojmañski 2008) or rotationally variable M dwarfs (Kiraga and
Stêpieñ 2007, hereafter KS2007).

In the present work we selected objects from the ROSAT Bright Source
Catalog (Voges \etal 1999) to have a good statistics of their X-ray
emission. Our primary goal was to obtain photometric periods for a large
sample of coronally active stars in order to study period--activity
relation. However, there are several classes of coronally active stars we
would like to study separately: pre-main sequence stars (various subclasses
of T~Tau type stars), single main sequence stars, binary main sequence
stars, evolved stars in binary systems (RS CVn type stars), single giants
(FK~Com type stars). In addition, coronally active stars are not the only
X-ray emitters, therefore there is a necessity to classify variable
objects: preliminary classification has been done on the basis of
photometric and X-ray properties and the data found in the literature. In
many cases catalogs of Strassmeier \etal (1993) or Eker \etal (2008) were
particularly useful.

This paper includes information concerning star selection and analysis
(Section~2), content of catalog (Section~3), overview of the data included
in the catalog and presentation of particularly interesting variable stars
(Section~4), summary and conclusions (Section~5).

The detailed period \vs X-ray activity relations for various carefully
chosen samples of coronally active variable stars will be given in the
forthcoming paper.

\Section{Star Selection and Analysis}
We based our work on the data collected by the ASAS (south) cameras located
at the Las Campanas, Chile. These observations, taken in {\it V}- and {\it
I}-bands, covered declinations below $+29\arcd$. We analyzed the {\it
I}-band data taken between March 25, 2002 and July 14, 2008, and the {\it
V}-band data taken between November 20, 2000 and October 27, 2008,
respectively.

The ROSAT All Sky Survey took place between July 30, 1990 and January 25,
1991 with some additional observations taken in February and August 1991
(Voges \etal 1999). Analysis of the data resulted in the detection of
145\,060 sources with 18\,811 sources classified as ``bright'' (ROSAT
all-sky survey bright source catalog -- RBSC, Voges \etal 1999). Bright
source has a count rate more than 0.05~cts/s in the 0.1--2.4~keV energy
band and has at least 15 photon counts.

There are 13\,793 entries from the RBSC south of declination $+29\arcd\!\!$.
The optical identification of the X-ray sources was based on coordinates
given in the ROSAT catalog and the search in the ASAS photometric database
was performed within 30\arcs around a given position (roughly two ASAS
pixels).

Several stars with large proper motions have not been identified in the
ASAS photometric database due to change of position between ROSAT and ASAS
epochs and have not been included in our present study. This is the case of
four rotationally variable M dwarfs from KS2007: GJ~84, GJ~176, GJ~205 and
GJ~551 (Proxima Centauri).

Good quality photometry is necessary for finding low amplitude variable
stars. The lower envelope of standard deviations for ASAS photometric
measurements is on the level of 0.01~mag for the {\it I}-band, 0.02~mag for
the {\it V}-band and increases steeply for stars fainter than 11~mag, to
reach 0.07~mag for a 12.5~mag stars in both bands. On the other hand stars
brighter than 8~mag are saturated.

We performed period search only for stars with the following properties:\\
(i)~the mean {\it I}- or {\it V}-band magnitude was between 8 and 12.5,
(ii)~for stars brighter than 8~mag dispersion was less than or equal to
0.1~mag, (iii)~number of data points was not less than 40.  These
conditions are similar to those adopted by KS2007.

The initial period search was performed for 5851 sources in the {\it
I}-band and and 4207 sources in the {\it V}-band (altogether 6026 stars)
using the AoV algorithm developed and described by Schwarzenberg-Czerny
(1989). Because we limit our search to low amplitude variability resulting
from spots, we decided to reject data points deviating from the mean
seasonal magnitude by more than 3.5 standard deviations, or by more than
0.35~mag. The average number of photometric measurements per star is about
225 for the {\it I}-band and 559 for the {\it V}-band.

Because many stars from our sample show long term variations, we have done
period search separately for each season, unless the number of useful
observations in the particular season was lower than 40 data points. A
similar period search was also performed on the whole data set after
subtracting season to season variations. For almost 3000 stars AoV
statistics was larger than 10 (using 6 phase bins); these stars were
further investigated using CLEAN algorithm (Roberts \etal 1987), and their
photometric data were inspected visually. Finally we obtained a list of
2302 objects with detected variability.

\subsection{Problems with Period Determination, Star Classification and 
Blending}
Variability of stars due to spots may not be strictly periodic what
introduces several problems. Formation of a new star-spot at the different
longitude than the previous one may result in a new minimum and phase shift
of the light curve. A new star-spot appearing at a different latitude on the
differentially rotating stellar surface results also in different period of
luminosity changes.

Objects monitored by ASAS are typically observed once per night or,
sometimes, even more sparsely, depending on the weather conditions. The
Nyquist frequency corresponds in this case to the period of 2~d. We looked
for periods from 0.1~d up to 200~d. Because periods of fast rotating stars
are shorter than the Nyquist limit, some obtained periods may be aliases of
the true ones. This problem is particularly difficult for low amplitude
variable stars with low signal to noise ratio. There are several examples
that the period considered as the best based on the ASAS photometry (using
AoV and CLEAN programs) is not correct but this can be checked only for the
well known stars like FK~Com, EZ~Peg, MQ~Lup, MS~Ser, or variable stars
found by surveys routinely observing the same field more frequently than
ASAS. For this reason we decided to compare our results to the work of
Norton \etal (2007, hereafter N07).

There are 428 variable stars found by SuperWASP project presented by N07.
We performed period search for 162 of these stars, finding 76 of them to be
variable. Forty object had periods similar to N07, 36 others had different
periods mostly due to aliases. In most cases we considered periods given by
N07 as more reliable than ours due to SuperWASP frequent sampling, so from
our aliased periods we chose those corresponding to the periods given by
N07. There were two cases where we had not adopted periods listed by N07.
We classified ASAS 065948+2742.0 as a low amplitude contact binary with a
period $P=0.328532$~d, whereas the period given by N07 is 0.1965~d (alias
to the half of our period). The star TYC 2064-1273-1 (ASAS 170313+2453.5)
is a rotational variable with a period of 11.57~d. During the SuperWASP
observations two groups of spots were present so the period given by N07 is
5.8417~d.

Uncertainties related to the presence of aliased periods for other stars
motivated us to search for additional data present in the literature. We
conducted the search mostly {\it via} SIMBAD database. As the most
important data we considered projected equatorial velocity ($\vv\sin i$),
lithium content and data on visual and spectroscopic companions.

Projected equatorial velocity together with a rotation period gives the
minimum radius of a star. Probability of a given inclination between the
rotation axis of a star and the observer's line of sight scales as
$\sin i$, so it is more probable to see a star with a mildly inclined
equator than almost pole on. Additionally, brightness variations due to
rotation of spotted surface are more difficult to measure for stars
observed almost pole on. Information about $\vv\sin i$ is usually
sufficient to discriminate between aliases indicating slow or fast rotation
(for example between aliased periods at $P=1.11$~d and $P=10$~d), however
we cannot use this method to discriminate between aliases that are close in
period domain (for example 2.22~d and 1.82~d).

There are numerous young stars on our list. These late type stars with
significant lithium abundance are still contracting toward the zero age
main sequence (ZAMS) and their dimensions may be significantly larger than
stars of the same spectral type already on the main sequence (MS). This
further limits the distinction between aliases based on the projected
equatorial velocity. Lithium line equivalent width is an important
indication of a stellar age. Lithium is also present in numerous giant
stars on our list, but lithium equivalent widths for these evolved stars
usually do not reach the level observed in very young stars.

For 520 stars (out of 2302) the periods are uncertain -- most significant
aliases were selected (see the Remark Section of the catalog).

Stellar classification is another important problem. When spots are stable,
resulting variability may resemble low amplitude close eclipsing binary,
ellipsoidal variability due to tidal deformation or pulsating star.
Stellar pulsations with significant amplitude are confined to well defined
regions in the Hertzsprung-Russel diagram, so using spectral types, or even
colors is often sufficient to exclude pulsation as a source of variability.
Distinction between close binaries and stars with stable star-spots is more
difficult. If amplitude and shape of light curve change, spots are
certainly present, but the presence of close companion star responsible for
period stability is also possible. Spectroscopic observations should
clarify ambiguity in classification of these stars.

In the case of detached binaries with large difference of surface
brightness we may face problem with detection of the secondary minimum
resulting in doubled period. There are several examples when a weak
secondary minimum is present only in the {\it I}-band data and for the {\it
V}-band data we observe only the main minimum.
 
Angular resolution of the ASAS cameras is similar to that of the ROSAT
survey. There were many cases when angular resolution of 15\arcs/pixel was
not enough to uniquely identify variable object. We inspected visually
Digitized Sky Survey (DSS) images of variable stars to evaluate degree of
blending. In the catalog we indicated stars with close neighbors found on
DSS frames with a letter ``B'' (stars of comparable brightness so it is
difficult to correctly identify the variable) or ``b'' (significantly
fainter stars close to the suspected variable; there is a small probability
that the fainter star was variable).

We also looked for close neighboring objects in available catalogs and
other publications. In this way we obtained information about visual binary
stars not resolved by DSS and about their separation and brightness
difference. We decided to include information about blended variable stars
together with the information about their known close neighbors. There are
1679 variable stars on our list without significant blending or known
visual neighbors.

\Section{Content of the Catalog}
The catalog includes information available in the SIMBAD database, ROSAT
bright source catalog (Voges \etal 1999), ASAS photometric data, and
some data taken from the literature.
\begin{itemize}
\parskip=0pt \itemsep=1mm \setlength{\itemsep}{0.4mm}
\setlength{\parindent}{-1em} \setlength{\itemindent}{-1em}
\item[]Column 1 -- ASAS designation equivalent to position from RBSC (${\rm hhmmss}\pm {\rm ddmm.m}$). 
\item[]Column 2 -- a distance between the formal ASAS position and related object from SIMBAD database (arc sec), note -- one pixel of ASAS detector $=14\arcs$.
\item[]Columns 3--12 include SIMBAD data (if present), the tilde ``$\sim$'' means no data available.
\item[]Columns 3--4 -- object name (two columns).
\item[]Column 5 -- object classification in the SIMBAD nomenclature.
\item[]Columns 6 and 7 -- proper motion in right ascension, and declination [mas/year].
\item[]Column 8 -- heliocentric parallax [mas].
\item[]Column 9 -- radial velocity [km/s].
\item[]Column 10 -- equivalent width of the lithium line at $\lambda{=}6704$~\AA\ expressed in {\AA}.
\item[]Column 11 --  projected rotational velocity ($\vv\sin i$~[km/s]):
\begin{itemize}
\setlength{\parindent}{-2em} \setlength{\itemindent}{-2em}
\parskip=0pt \itemsep=1mm \setlength{\itemsep}{0.4mm}
\item[] ``a'' -- there are substantial differences in $\vv\sin i$ values found in literature,
\item[] ``b'' -- data about $\vv\sin i$ are probably related to a companion star,
\item[] ``$<$'' -- the upper, and  ``$>$'' -- the  lower limit for $\vv\sin i$. 
\end{itemize}
\item[]Column 12  visual companions (more information in remarks):
\begin{itemize}
\setlength{\parindent}{-2em} \setlength{\itemindent}{-2em}
\parskip=0pt \itemsep=1mm \setlength{\itemsep}{0.4mm}
\item[] ``-'' --  no known visual companion,
\item[] b -- blended with much fainter star or stars,
\item[] B -- blended with a star or stars of comparable brightness,
\item[] (b and B -- information based on visual inspection of DSS frames -- usually DSS-2-red),
\item[] c -- close visual companion (visual binary) much fainter than the primary star,
\item[] C -- close visual companion of comparable brightness,
\item[] (c and C -- information based on SIMBAD database or literature - references in  ``remarks'').
\end{itemize}
\item[] Column 13 -- close or spectroscopic companions (more information in ``remarks''):
\begin{itemize}
\parskip=0pt \itemsep=1mm \setlength{\itemsep}{0.4mm}
\setlength{\parindent}{-2em} \setlength{\itemindent}{-2em}
\item[] p -- photometry may indicate a close companion,
\item[] P -- definitely eclipsing binary (based on photometric data),
\item[] SB1, SB2 -- spectroscopic binary (single lined and double lined, respectively),
\item[] SB3, SB4 --  triple and quadruple spectroscopic systems,
\item[] ? -- there is no data about spectroscopic variability (no radial velocity measurement or single radial velocity measurement, no spectroscopic lines of secondary star) and there is no photometric indication of close companion,
\item[] no -- radial velocity measurements indicate constant radial velocity (no close companion star detected),
\item[] C -- composite spectrum of two or more stars, but without noticeable changes in the radial velocity.  
\end{itemize}
\item[] Columns 14 and 15  -- {\it B} and {\it V} magnitude (SIMBAD database).
\item[] Column 16 --  spectral type. When no data available we put a tilde ``$\sim$``, references to particular measurements not listed  in the SIMBAD database are listed in remarks.
\item[] In the columns 17 -- 22  we include the basic ASAS photometric data:
\begin{itemize}
\setlength{\parindent}{-2em} \setlength{\itemindent}{-2em}
\parskip=0pt \itemsep=1mm \setlength{\itemsep}{0.4mm}
\item[] Column 17 --  number of observations in the {\it I}-band,
\item[] Column 18 -- mean magnitude in the {\it I}-band,
\item[] Column 19 -- dispersion of the {\it I}-band measurements [mag], 
\item[] Column 20 --  number of observations in the {\it V}-band,
\item[] Column 21 -- mean magnitude in the {\it V}-band,
\item[] Column 22 -- dispersion of the {\it V}-band measurements [mag].
\end{itemize}
\item[] Column 23 -- adopted bolometric correction for the {\it I}-band ($BC_I$), based on $(V-I)$ color (calculated from mean {\it I}- and {\it V}-band magnitudes) using a fit to atmospheric models presented by Bessel \etal (1998) in the form $BC_I=-0.08+1.86(V-I)-1.33(V-I)^2+0.251(V-I)^3$ for values of $(V-I)$ in the range of $(0{-}1.5)$, and a slightly changed formula given by Reid and Gilmore (1984) for  values of  $(V-I)$ in the range of $(1.5{-}4.6)$: $BC_I=-0.31(V-I)+1.03 $. The value of the free parameter is changed from 1.05 to 1.03 to obtain continuity with the formula for smaller values of $(V-I)$. There are 5 stars without {\it V}-band measurements and 6 stars without {\it I}-band  measurements. We have not calculated bolometric corrections for these stars.
\item[] Columns 24--27 include some observational data from RBSC: 
\begin{itemize}
\setlength{\parindent}{-2em} \setlength{\itemindent}{-2em}
\parskip=0pt \itemsep=1mm \setlength{\itemsep}{0.4mm}
\item[] Columns 24 and 25 --  number of counts per second and its error,
\item[] Columns 26 and 27 --  hardness ratio {\it HR1}, and its error $(HR1=(B-A)/(A+B)$, where $A$ is a number of counts in energy band $0.1{-}0.4$~keV, and $B$ -- a number of counts in energy band $0.5{-}2.0$~keV).
\end{itemize}
\item[] Column 28 and 29 -- logarithm (base 10) of the X-ray to the bolometric flux ratio and its minimal formal error (based only on uncertainties of the fit of the coronal model to X-ray observations). A conversion from the count ratio and {\it HR1} index to the flux in the $0.1{-}2.4$~keV photon energy range is given by Schmitt \etal (1995):~~ $F_x=(5.31{\it HR1}+8.31)\cdot10^{-12}~{\rm cts}.$ Here $F_x$ is expressed in ergs/cm$^2$/s and ''cts'' means counts per second listed in the ROSAT catalog. This formula was derived for X-ray coronal emission, but we adopted it for all stars. The value of $R_x$ is calculated as $F_x/F_{\rm bol}$ where $F_{\rm bol}$ was obtained from mean {\it I} magnitude, bolometric correction $BC_I$, and the assumption that the solar absolute bolometric magnitude is equal to 4.75.
\item[] Column 30  -- adopted period of luminosity changes. 
\item[] Column 31 -- lower limit for maximum amplitude of the {\it I}-band  variability [mag].
\item[] Column 32 -- lower limit for maximum amplitude of the {\it V}-band variability [mag]. Amplitudes of some variable stars (especially eclipsing variables) may be underestimated due to the rejection of extreme data points.
\item[] Column 33 -- information about the photometric variability type:
\begin{itemize}
\setlength{\parindent}{-2em} \setlength{\itemindent}{-2em}
\parskip=0pt \itemsep=1mm \setlength{\itemsep}{0.4mm}
\item[] ED -- detached eclipsing binary,
\item[] EB -- close eclipsing binary with deformed component, and unequal depths of eclipses,
\item[] EC -- contact eclipsing binary,
\item[] Ell -- variability due to the deformation of a star in a close binary system,
\item[] rot -- rotational variability due to the presence of spots, 
\item[] puls -- variability due to stellar pulsations,
\item[] msc - the variability is difficult to interpret.
\end{itemize}
\end{itemize}

There are 1936 stars (likely 1236 new) listed as variable due to the
presence of spots (rot). Our list also includes 127 detached eclipsing
binary stars (33 new), 124 contact binaries (11 new), 96 eclipsing stars
with deformed components (19 new), 13 ellipsoidal variable stars (4 new), 5
miscellaneous variables and one pulsating RR~Lyr type star (blended with an
eclipsing binary).

After column 33 we put our own remarks and data from the literature about
particular objects. Information from different sources is separated by
semicolons and may include:
\begin{itemize}
\parskip=0pt \itemsep=1mm \setlength{\itemsep}{0.4mm}
\item[] P\_ACVS -- period from  ASAS Catalog of Variable Stars,
\item[] Pphot -- photometric period found in the literature (other sources then ACVS),
\item[] n(vrad) -- number of radial velocity measurements in cited paper,
\item[] sig(vrad) -- dispersion of radial velocity measurements given in cited paper,
\item[] Porb -- spectroscopic orbital period (with K1 and K2 radial velocity semi-amplitudes when available),
\item[] RS -- RS CVn type variable star -- coronally active evolved star in a close binary system,
\item[] RS? -- probably RS CVn type star (based on photometric behavior and X-ray data), but its binarity should be confirmed,
\item[] EW Li line -- equivalent width of Li line at 6704~{\AA},
\item[] vis bin -- visual binary star with a given separation and luminosity contrast given by the difference in magnitudes at given band, or measured flux ratio at given band.
\item[] F03 -- star is listed in the catalog of Fuhrmeister and Schmitt (2003) and has variable X-ray emission 
\end{itemize}

\renewcommand{\arraystretch}{1.2}
\MakeTableee{cccccccc}{12.5cm}{The catalog of variable stars}
{\hline
col. 1          & col. 2  & col. 3      & col. 4      &  col. 5       & col. 6         & col. 7         & col. 8\\ \hline
ASAS des        & sep     & object      & name        & SIMBAD type   & $\mu_{\alpha}$ & $\mu_{\delta}$ & plx\\ \hline
000724-4233.4   & 12.55   & HD          & 271         & *             & $-14.80$       & $ -4.00$       &  $\sim$\\
000921+0038.1   & 11.86   & TYC         & 1-1187-1    & pr*           & $119.40$       & $-29.70$       &  $\sim$\\ 
001009-5921.2   & 15.97   & CD-60       & 8           & *i*           & $ 15.40$       & $  6.10$       &  $\sim$\\ \hline
col. 9          & col. 10 & col. 11     & col. 12     & col. 13       & col. 14        & col. 15        & col. 16\\ \hline
$\vv_{\rm rad}$ & Li EW   & $\vv\sin i$ & blend       & sp. comp.     & $B$            & $V$            & sp. type\\ \hline 
$-56.4$         & $\sim$  & $\sim$      &    --       & p             &   9.97         &  9.55          & F5/F6V\\
$-20  $         & 0.10    & 28.1        &    --       & SB2           &  12.20         & 11.88          & K4Ve\\ 
$-10  $         & 0.11    & 18.0        &     B       & SB2           &  10.78         & 9.92           & G8III \\ \hline
col. 17         & col. 18 & col. 19     & col. 20     & col. 21       & col. 22        & col. 23        & col. 24\\ \hline
n(I)            & \meani  & $\sigma(I)$ & n(V)        & \meanv        & $\sigma (V)$   & BC(I)          & cts\\ \hline
138             &  8.881  & 0.034       & 424         &  9.482        & 0.034          & 0.61           & 0.087\\
111             & 10.252  & 0.043       & 281         & 11.452        & 0.060          & 0.67           & 0.114\\
167             &  8.747  & 0.033       & 596         &  9.664        & 0.069          & 0.70           & 0.163\\ \hline
col. 25         & col. 26 & col. 27     & col. 28     & col 29        & col 30         & col 31         & col 32\\ \hline
e(cts)          &{\it HR1}& e(HR1)      & $\log(R_x)$ & e($\log(R_x)$)& $P$            & amp($I$)       & amp($V$)\\ \hline
0.025           & $ 0.10$ & 0.29        & $-3.71$     & 0.46          &  0.8391        & 0.052          & 0.061\\
0.019           & $-0.01$ & 0.16        & $-3.05$     & 0.27          &  1.6419        & 0.060          & 0.069\\
0.036           & $ 0.78$ & 0.18        & $-3.31$     & 0.30          & 23.56          & 0.033          & 0.102\\ \hline
col. 33         &         &             &             &               &                &                &\\ \hline
type            & \multicolumn{7}{c}{Remarks}\\ \hline
 rot            & \multicolumn{7}{l}{also possible Ell or EB and $P=1.6782$~d; n(vrad)=1 (Zw2008)}\\   
 rot            & \multicolumn{7}{l}{SB2, $\vv\sin i=28.1$~km/s, n(vrad)=1, EW Li line = 0.1 A (T06); P\_ACVS 0.621853;}\\
 rot            & \multicolumn{7}{l}{RS; SB2, dV=2.5, $\vv\sin i=18$~km/s, n(vrad)=1, EW Li line = 0.11 A (T06);}\\
}

\begin{figure}[p]
\vglue-3mm
\includegraphics[width=14cm]{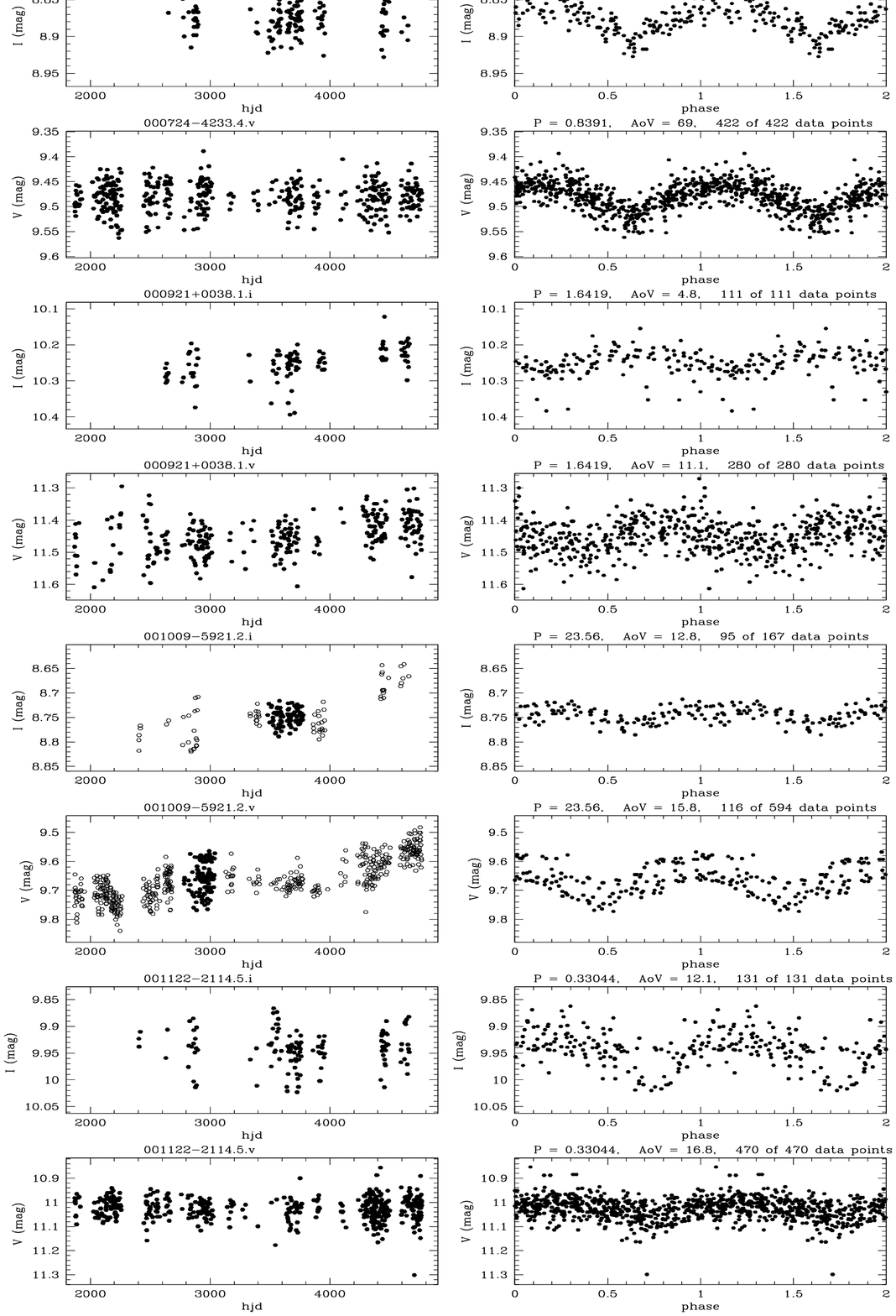}
\FigCap{Sample of stellar photometric data given as 
a function of ${\rm hjd=HJD-2450000}$~d, ({\it left panels}) and phased
periodic variability ({\it right panels}). Figures for all stars are
available in the electronic form from the {\it Acta Astronomica Archive} ({\it
ftp://ftp.astrouw.edu.pl/acta/2012/kir\_67}).}
\end{figure}
References to most often cited papers are written in the abbreviated
form:\\ 
Al2000 -- Alcala \etal (2000), Bal87 -- Balona (1987), Ber09 -- Bernhard
\etal(2009), Cov97 -- Covino \etal (1997), CCDM -- ``Catalogue of the
Components of Double and Multiple Stars'' -- Dommanget and Nys (2002),
Cut99 -- Cutispoto \etal (1999), daSil09 -- da Silva \etal (2009), Guen07
-- Guenther \etal (2007), Guil09 -- Guillout \etal (2009), GCVS --
``General Catalogue of Variable Stars'' - Samus \etal (2009), KS2007
-- Kiraga and Stêpieñ (2007), KE2002 -- Koen and Eyer (2002), LH98 -- Li
and Hu (1998), LEK87 -- Lloyd Evans and Koen (1987), Nord04 -- Nordstrom
\etal (2004), N07 -- Norton \etal (2007), R06 -- Riaz \etal (2006), Str2000
-- Strassmeier \etal (2000), T06 -- Torres \etal (2006), WGH07 -- White
\etal (2007), W98 -- Wichmann \etal (1998), W99 -- Wichmann \etal (1999),
W2000 -- Wichmann \etal (2000).  Zw2008 -- Zwitter \etal (2008),

In the remarks we included also uncertainties about classification of the
star (usually whether it is a low amplitude contact eclipsing binary or
deformed binary star or rotationally variable spotted star), or its period
(usually we listed the most significant aliases). Table~1 presents a sample
of our catalog. Full version is available in the electronic form from the
{\it Acta Astronomica Archive}\footnote{\it
ftp://ftp.astrouw.edu.pl/acta/2012/kir\_67}. Files containing figures with
photometric data and phased light curves of all variable stars
(VAR\_00\_04.ps, VAR\_04\_05.ps, VAR\_05\_06.ps, VAR\_06\_07.ps,
VAR\_07\_09.ps, VAR\_09\_12.ps, VAR\_12\_14.ps, VAR\_14\_16.ps,
VAR\_16\_17.ps, VAR\_17\_18.ps, VAR\_18\_21.ps, VAR\_21\_24.ps) are also
available for download from {\it Acta Astronomica Archive}. Stars are sorted by
increasing right ascension, and the file VAR00\_04.ps contains variables in
right ascension range [0h, 4h).  The first page of the file VAR\_00\_04.ps
is presented in Fig.~1.  References to the literature cited in the catalog
are in the file ``refs.dat''.

\Section{The Overview of the Catalog Content}
Our catalog contains 2302 variable stars. The discussion of statistical
properties of the sample and a presentation of interesting objects is given
below.

\subsection{The V and I magnitude and $(V-I)$ color distribution}
Histogram illustrating the number of stars in 0.5~mag bins is presented in
the left panel of Fig.~2 ({\it I}-band -- the dotted line, {\it V}-band --
the solid line) and has maximum around $I=9.5$~mag and $V=10.5$~mag. There
are variable stars down to the $V=14.5$~mag, but they have been found to be
variable using only {\it I}-band data. As we expect for population of stars
thought to be dominated by coronal sources the most of variable stars have
$(V-I)$ color typical for stars with outer convection zones (see the right
panel of Fig.~2). The maximum around $(V-I)=1.0$~mag indicates that K-type
stars are most numerous in our sample.  There are six stars with
$(V-I)>3$~mag, with $(V-I)=3.46$~mag for the reddest one (ASAS designation
173353+1655.2). A few stars with the bluest colors $(V-I)<0.3$~mag are not
expected to be coronal sources. In this color range we expect main sequence
stars of early spectral type (not later than A3) and cataclysmic
binaries. This is confirmed by spectral classification found in the SIMBAD
database.
\begin{figure}[htb]
\vspace*{-6pt}
\includegraphics[width=12.2cm, bb=23 423 573 693]{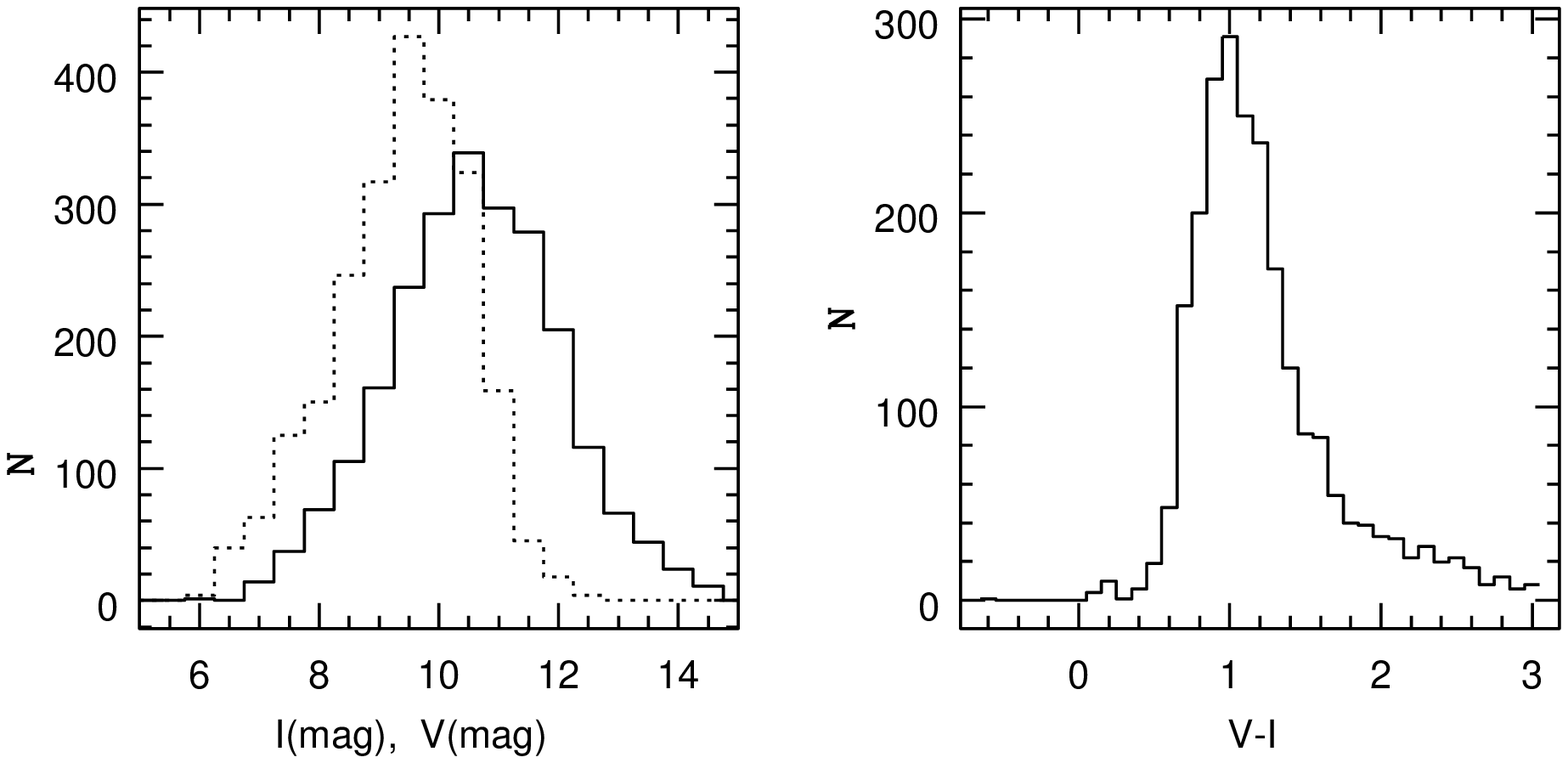}
\vspace*{-4pt}
\FigCap{Basic photometric properties of 2302 periodic variable stars 
illustrated in histograms. {\it Left panel}: number of stars in 0.5~mag
bins: {\it I}-band sources -- dotted line, {\it V}-band sources -- solid
line. {\it Right panel}: number of stars as a function of $V-I$ color
(0.1~mag bin in $V-I$).}
\end{figure}

Some stars close to the Galactic plane or star formation regions may be subject
to significant absorption and reddening. {\it E.g.,} NGC 3603 A1 (Moffat
\etal 2004) and HT Lup (Prato \etal 2003) have total absorption of 4~mag in
the {\it V}-band. For most of the stars we have no data about their
reddening and we could use maps of total Galactic extinction (Schlegel
\etal 1998) as an indication of maximum extinction in a given direction.
 
We do not have $(V-I)$ color data for 10 stars, of which 5 stars are
without {\it V}-band measurements and 6 stars are without {\it I}-band
measurements.

\subsection{Spectral Classification} 
Spectral types, taken mainly from the SIMBAD database, are given for 1699
stars and are summarized in Table~2 (7 stars with luminosity class IV/III
are included in the III luminosity class, and 10 stars IV/V are in the IV
luminosity class). There are also known 11 white dwarfs related with stars
listed in the present catalog, six are components of cataclysmic binaries
(BL~Psc, IX~Vel, DH~Leo, BV~Cen, AE~Aqr, and most likely V341~Ara), and
five are close neighbors of variable stars. Probably there is also a
neutron star present in the X-ray symbiotic star IGR~J16194-2810.

\MakeTable{|c|c|c|c|c|c|c|c|c|}{12.5cm}{Spectral classification 
of the primary stars listed in the catalog}
{\hline
spectral type    & O &  B &  A &  F &   G  &   K  &   M & tot \\
luminosity class &   &    &    &    &      &      &     &     \\ 
\hline 
 II        & 0 &  0 &  0 &  0 &   0 &   2  &   0 &   2 \\
 III       & 0 &  0 &  0 &  3 &  84 & 138  &   4 & 229 \\ 
 IV        & 0 &  0 &  3 & 10 &  81 &  90  &   1 & 185 \\
 V         & 1 &  1 &  5 & 67 & 225 & 353  &  82 & 734 \\
 not given & 2 &  3 & 15 & 44 & 175 & 235  &  75 & 549 \\
\hline
total      & 3 &  4 & 23 &124 & 565 & 818  & 162 & 1699\\ 
\hline}

Most stars have spectroscopic classification typical for the coronally
active stars. For early type O and B stars we expect X-ray emission from
binary stars as a result of colliding winds, or interactions in close
systems.

Among three O-type stars there are two eclipsing binaries, NGC3603 A1
(111510-6115.6) and MY~Ser (181806-1214.6), and one cataclysmic variable,
V341~Ara.

Among four B-type stars, two, HD~79905 (091407-6032.0) and HD~195134
(202908+1241.0), are small amplitude (${\rm d}V\approx0.05$~mag) eclipsing
or ellipsoidal variable stars, one, LM~Lup, is a rotationally variable,
chemically peculiar $\alpha$~CVn type star, and one, IX~Vel
(081519-4913.2), is a cataclysmic variable star.

There are only two stars of luminosity class II in our catalog: MQ~Vir
(135952-2220.8), described later, and CD-46 10694 (162335-4631.7) with a
period of 44.7~d. The latter star is a fast rotating giant with a projected
equatorial velocity equal to 35~km/s (T06). Assumption that variability
period is equal to the rotation period gives the minimal radius of the star
30 times larger than the solar radius. This requires the presence of a
companion to tidally spin up the giant star to a present rotation rate.

Late type giants are known not to be coronally active (G\"udel 2004). This
is also apparent in our study where we find four stars classified as M-type
giants but only for two of them X-ray emission may be directly related to
the giant star itself.

The star GSC 06806-00016 (IGR J16194-2810, ASAS 161933-2807.5) is
classified as a symbiotic X-ray binary and its X-ray emission is related to
the compact companion. The M2III star HD 184189 is blended with a fainter
companion ($\Delta V=1$~mag) so eclipses with a period of $P= 2.1334$~d
(and most probably X-ray emission too) are not related to the red giant
star. The M0IIIe type star ASAS 092505-4327.8, is very unusual with the
high amplitude variability (up to 0.5~mag in the {\it V}-band), fast
rotation ($\vv\sin i=35$~km/s) and the high activity
($\log(R_x)=-2.6$). Most probably it is a very young star still contracting
toward the main sequence ($R_{\rm min}=7~\RS$) rather than the evolved one.

The only variable M type giant with X-ray emission and a firm spectroscopic
classification is KP Aqr (unresolved Hipparcos variable). We adopted 21.8~d
as variability period for this star, but other periods were also
detected. The rotational variability cannot be reconciled with star's
parameters. The recent parallax calculation for KP~Aqr, based on Hipparcos
data (van Leeuwen 2007), gives a distance of about 400~pc, so its
bolometric luminosity is about 400~\LS\ and its radius is more than
40~\RS. Rotation with a period equal to 22~d would be faster than the
breakup rotation for a solar mass star with $R=40$~\RS. The variability may
be related to small amplitude pulsations (like OGLE Small Amplitude Red
Giant stars, Soszyñski \etal 2011), but this suggestion needs to be
confirmed.

Small number of M giants in our sample indicates that most of M-type stars
with not given luminosity class are dwarfs.

\subsection{Period Distribution}
We limited our search to the periods between 0.1~d and 200~d. Periods range
from 0.1371~d (for the ASAS 102045-6311.3) up to 192.8~d (IGR J16194-2810=
ASAS 161933-2807.5). Their distribution presented in Fig.~3 has maximum
around 3~d. Most stars (95.5\%) have periods between 0.29~d and 50~d.
There are five stars in our catalog with the periods above 100~d:
083531-0904.1 ($P=171.2$~d), NV~Hya (091701-0937.1, $P=145$~d), MQ~Vir
(135952-2220.8, $P=120.4$~d), GSC 06806-00016 (161933-2807.5, optical
counterpart to the gamma ray source IGR J16194-2810, $P=192.8$~d), and HD
341626 (180816+2350.5, $P=123.4$~d). For those stars photometric
observations and phased light curves are presented in Fig.~4. Three of
them, already known to be variable (NV~Hya -- Perryman \etal 1997, MQ~Vir
-- Koen and Eyer 2002, HD 341626 -- ACVS), are K0-type giants ($1.1<(V-I)<
1.3$), characterized by rather hard X-ray emission (${\it HR1}=0.73{-}1.0$)
and quite high activity ($\log(R_x)\approx-4$). Most probably, they are
active coronal sources. None of these stars is confirmed to be a
spectroscopic binary, but they are candidates for RS~CVn type stars due to
their spectral types, character of photometric variability and the high
X-ray activity.  The star, ASAS 083531-0904.1, is an optical counterpart to
the X-ray source 1RXS J083531.6-090409. Its variability period is close to
170~d, but there are also possible aliases close to 1~d or 0.5~d. According
to Schlegel
\etal (1998) Galactic extinction in the direction of this star is moderate
($A_V=0.144$~mag) so reddening is not significant. The red color of the star
($V-I=1.9$~mag) and relatively high activity ($\log(R_x)=-3.4$) may indicate
that it is a red dwarf, or that its X-ray emission is of non coronal
origin.
\begin{figure}[p]
\centerline{\includegraphics[width=7cm]{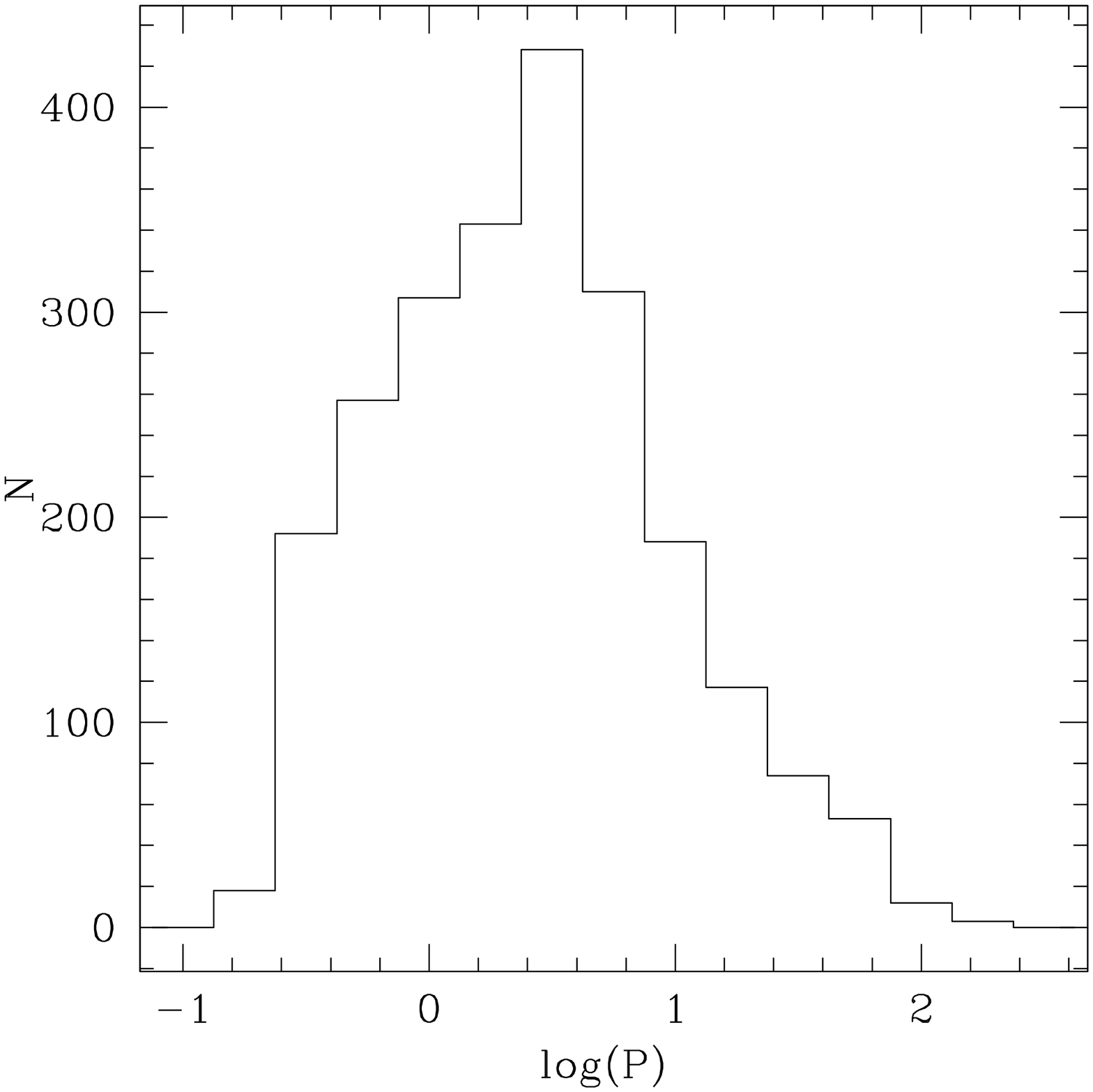}}
\FigCap{Histogram of variability periods. Number of stars in $\log(P)$ bins
equal to 0.25 is (period is expressed in days) is given.}
\vskip5mm
\centerline{\includegraphics[width=13cm, bb=23 250 575 705]{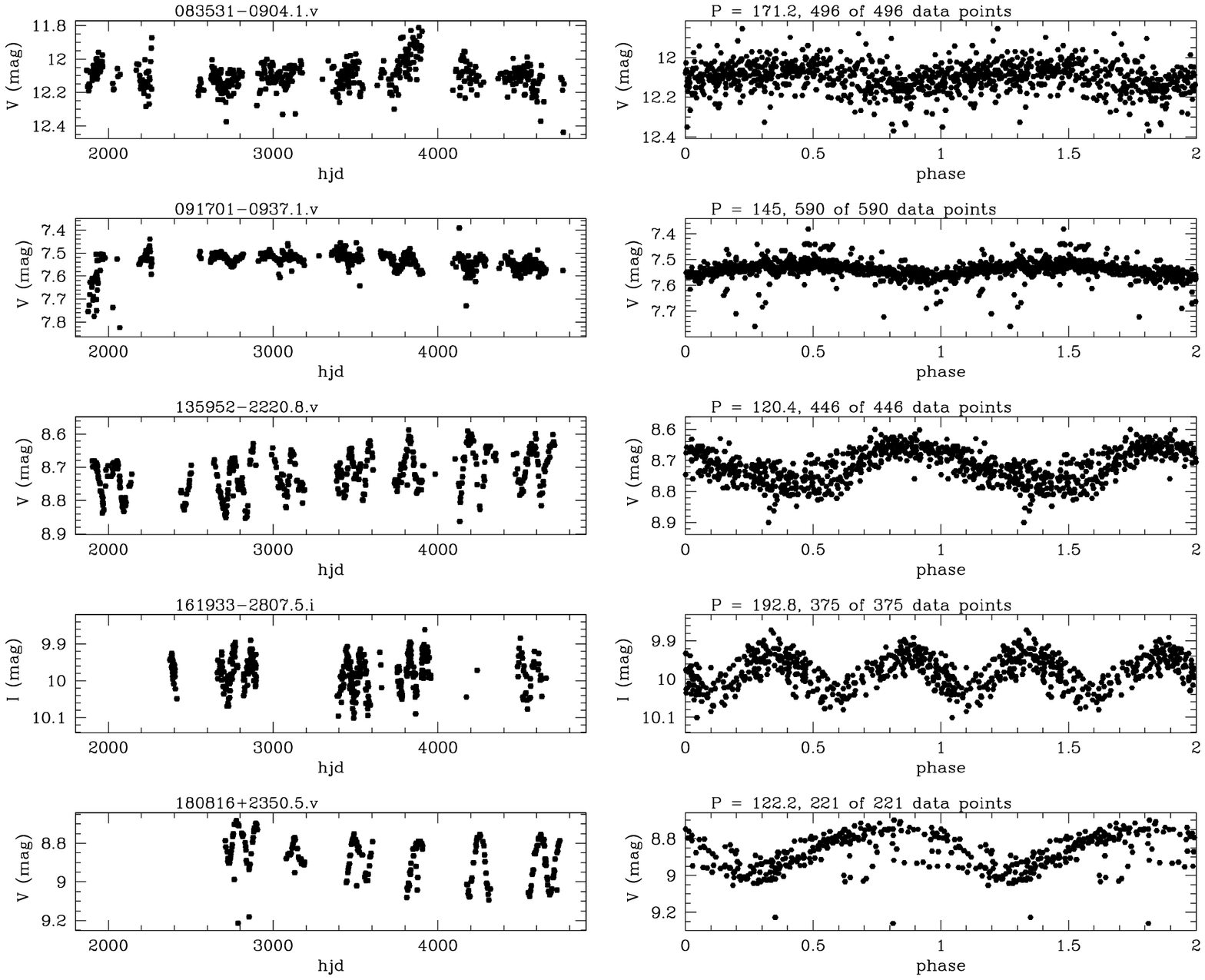}}
\FigCap{Stars with the longest variability periods.
{\it Left panels:} photometric data as a function of ${\rm
hjd=HJD-2450000}$~d. {\it Right panels}: photometric data points phased
with the adopted variability period.}
\end{figure}
 
GSC 06806-00016 (ASAS 161933-2807.5) is an optical counterpart to the gam\-ma
and X-ray source (IGR J16194-2810) classified as a symbiotic X-ray binary
(Masetti \etal 2007). The X-ray source coincides with the red giant of
spectral type M2III but very red color $(V-I)=2.75$~mag, results most
probably from the interstellar reddening ($A_V=3.2$~mag in the direction to
the star, Schlegel \etal 1998). We detected {\it I}-band photometric
variability with an amplitude of 0.11~mag and a period of 192.8~d what we
interpret as a result of ellipsoidal distortion due to the presence of a
massive companion. However, the pulsation with the period two times
shorter is also possible. The object is very active ($\log(R_x)=-2.94$)
and has been detected also above 30~keV by INTEGRAL (its ROSAT {\it HR1}
``color'' is equal to 0.96). The star is a LMXB with the compact object
(neutron star or black hole) and the giant star companion.

\begin{figure}[htb]
\centerline{\includegraphics[width=13cm]{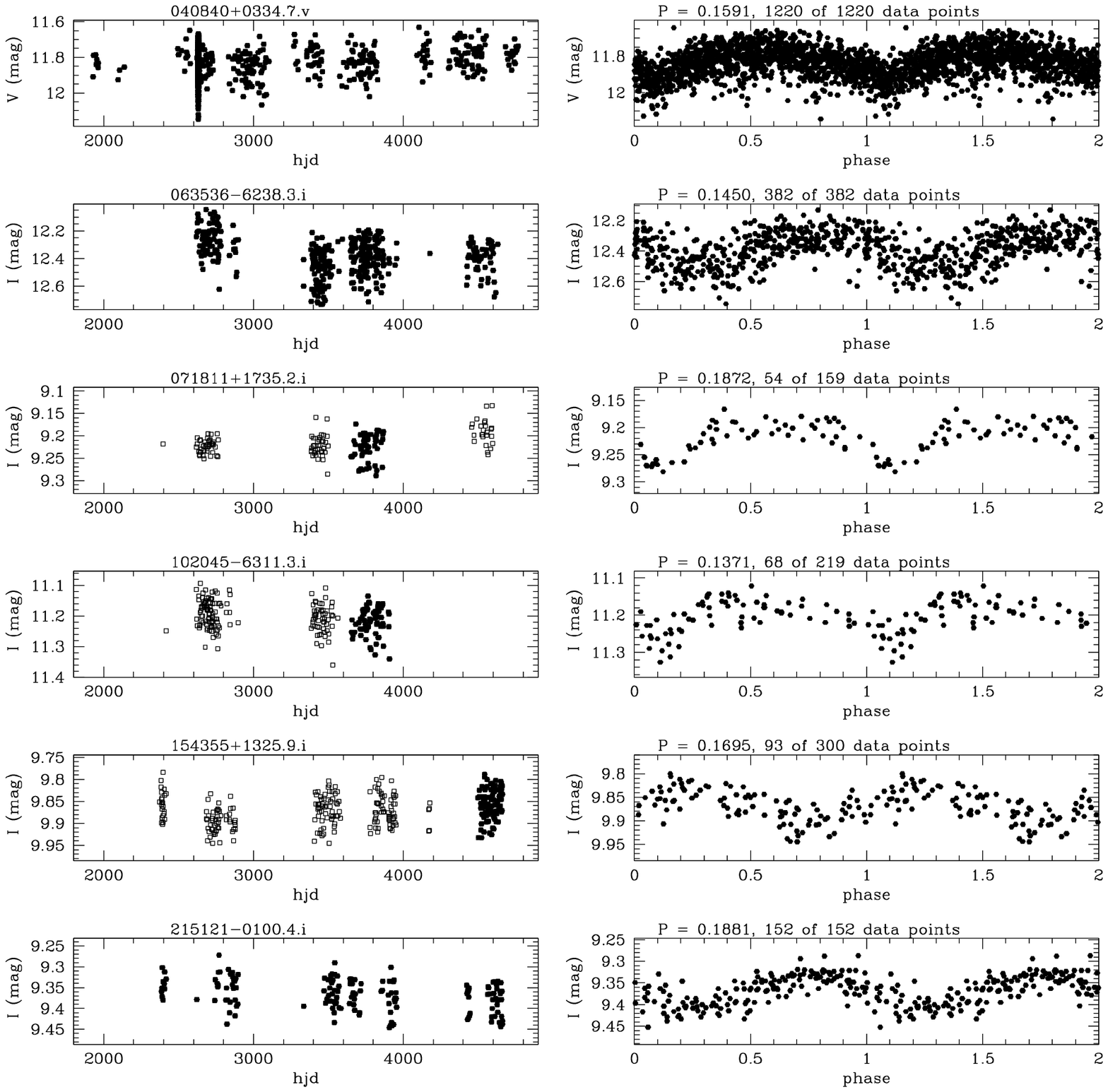}}
\vspace*{-7mm}
\FigCap{Stars with the shortest variability periods.
{\it Left panels}: photometric data as a function of ${\rm hjd=HJD-
2450000}$~d. {\it Right panels}: photometric data points phased with the
adopted variability period.}
\end{figure}
Stars with rotation periods shorter than 0.2~d are usually considered as a
very fast rotators. They are tracers of the apparent decline in activity
called supersaturation (Prosser \etal 1996). There are six stars with
periods shorter than 0.2~d in our sample. One of the stars is RR~Pic (Nova
Pic 1925, ASAS 063536-6238.3), well known cataclysmic variable star which
erupted as a slow nova in 1925. The orbital period of $P=0.1450255$~d was
determined by Vogt (1975). The star has a blue color ($I=12.39$~mag,
$V=12.50$~mag) and its X-ray emission is most probably of non coronal
origin. Other stars with periods shorter than 0.2~d are probably coronal
sources of spectral type K (ASAS 071811+1735.2, 154355+1325.9,
215121-0100.4) and M (ASAS 040840+0334.7, 102045-6311.3). Two of these
stars are known to be variable (ASAS 154355+1325.9 with $P=0.16952$~d, ASAS
215121-0100.4 with $P=0.18816$~d, ACVS) and our periods are essentially the
same. Photometric observations and phased light curves of six stars with
shortest periods in our sample are presented in Fig.~5.

ASAS 102045-6311.3 corresponds to the X-ray source 1RXS J102045.8-631121
and has the shortest period ($P=0.1371$~d) on our list. The identification
of the variable star is not certain because of blending. The brighter star
is red, $(V-I)=2.2$~mag, and has spectral type M4 (Riaz \etal 2006). Its
very high activity ($\log(R_x)=-2.4$ and ${\it HR1}=0$) may confirm that
for very late type dwarfs supersaturation is not observed (Jeffries \etal
2011).

\subsection{Variability Amplitude of Spotted Stars}
The largest amplitude of variability related to the presence of star-spots
has been observed for stars above the main sequence. The best known
examples are the pre-main sequence star V410 Tau with $\Delta V$ up to
0.65~mag (Strassmeier \etal 1997) and the evolved (spectral type K0III)
RS~CVn type star XX~Tri with $\Delta V$ up to 0.63~mag (Strassmeier 1999,
Strassmeier 2009).

Most of the rotationally variable stars found in our study have small
amplitudes as can be seen in the histogram presented in Fig.~6. More than
70\% of the newly detected variable stars have the amplitudes smaller than
0.1~mag and only 47 among them have the amplitudes larger than 0.2~mag.

\begin{figure}[htb]
\centerline{\includegraphics[width=7cm]{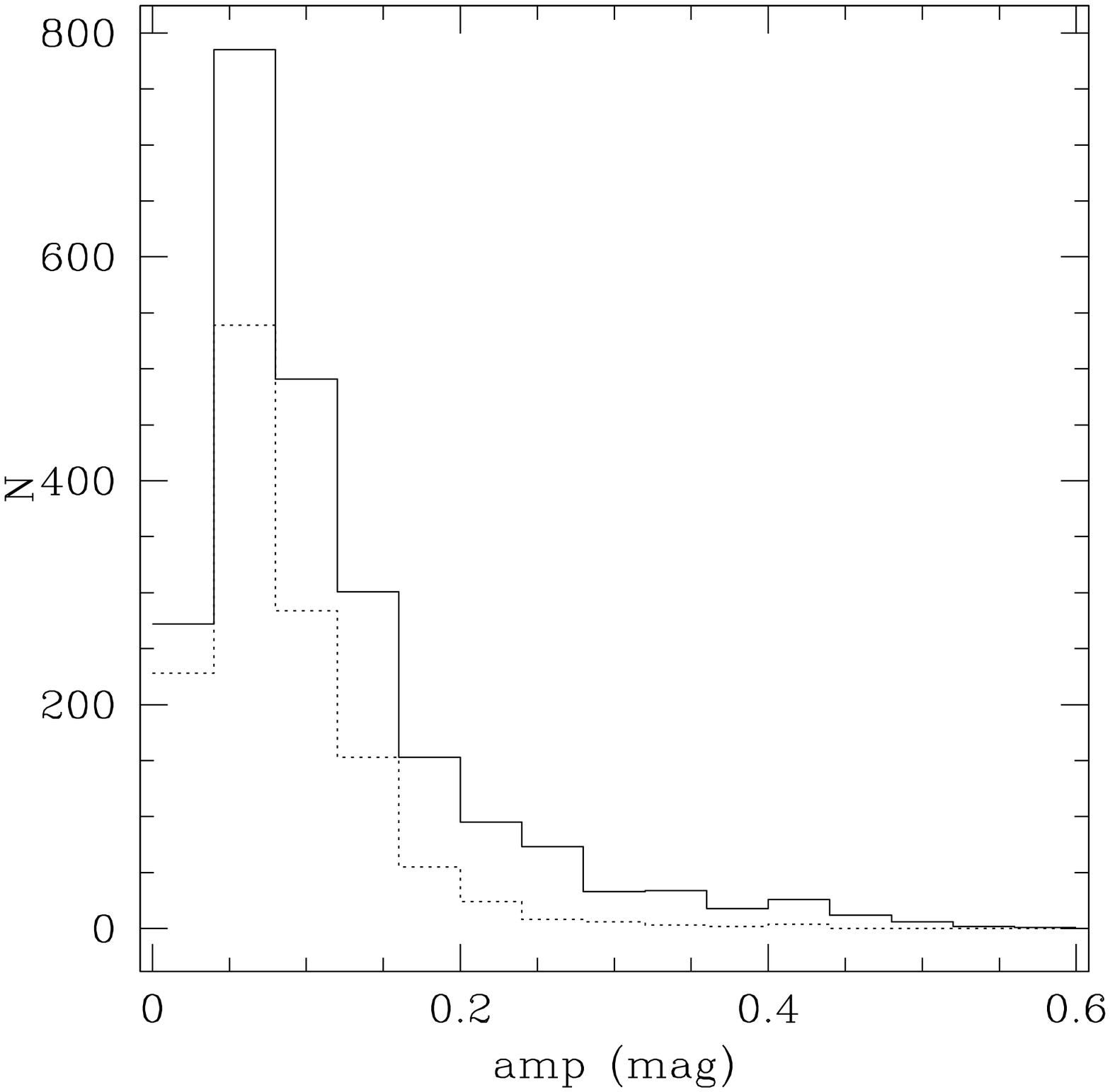}}
\FigCap{Histogram of maximum variability amplitude of variable stars.
Number of stars in bins of variability amplitude is equal to 0.04~mag. All
periodic variable stars -- solid line, new periodic variable stars --
dotted line.}
\end{figure}
\begin{figure}[htb]
\centerline{\includegraphics[width=13cm, bb=23 270 575 700]{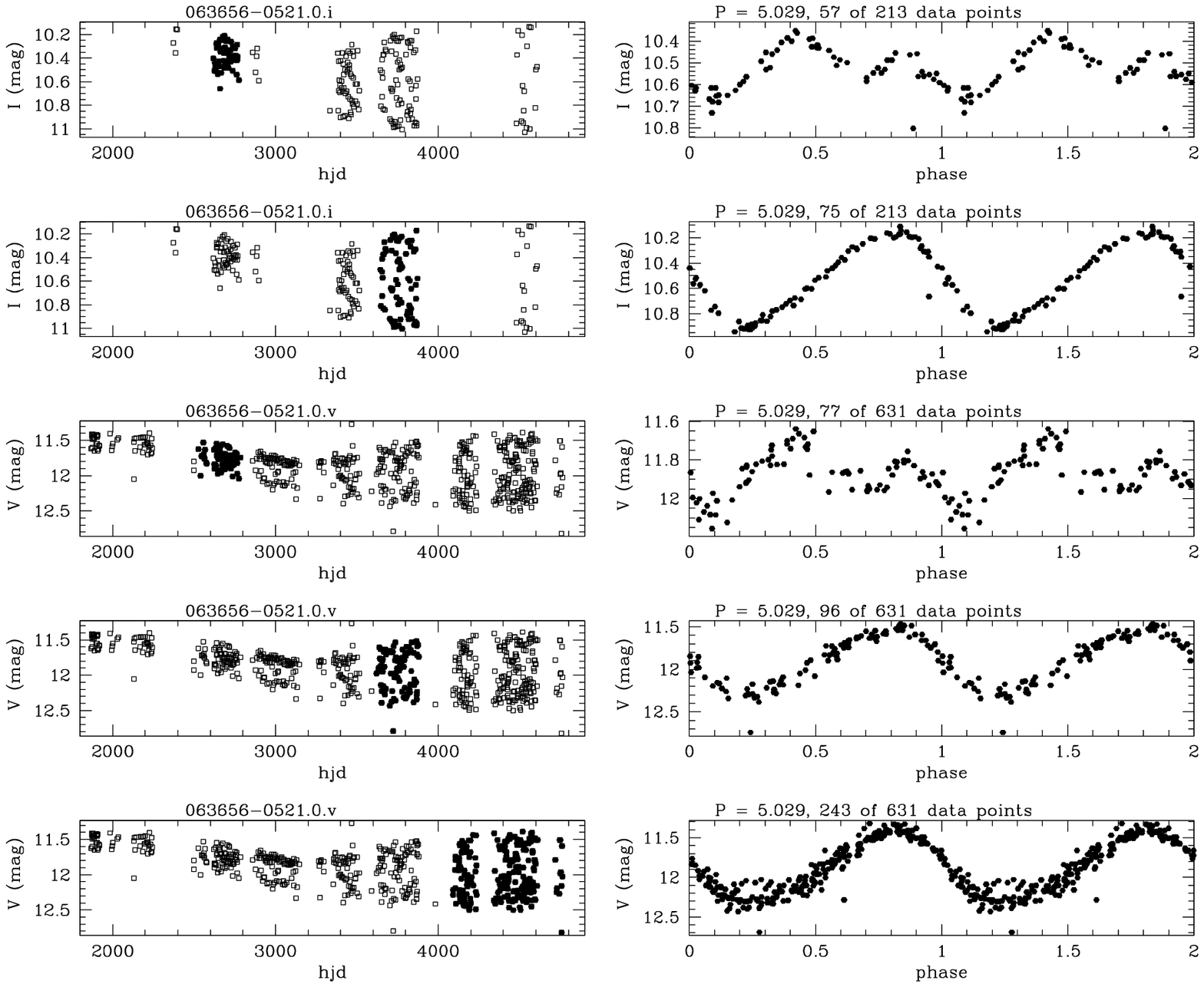}}
\FigCap{ASAS 063656-0521.0 photometric data ({\it left panels}) 
and the phased light curve ({\it right panels}). The {\it I}-band
light curve is given for two different seasons ({\it two upper rows}) and
{\it V}-band light curve is given for three different seasons ({\it three
lower rows}).}
\end{figure}
One of the new variables, ASAS 063656-0521.0, presents particularly large
luminosity changes due to presence of star-spots.  It is an optical
counterpart to the X-ray source 1RXS J063656.7-052104. We classified this
star as a rotational variable with a highest seasonal brightness
variability (up to $\Delta V=0.8$~mag). This star showed significant
activity changes. Its mean {\it V} dropped from 11.5~mag to 12.0~mag
between ${\rm hjd=1880}$ and ${\rm hjd=4200}$ (with $V-I$ color changing
from 1.26~mag to 1.38~mag). At the same time variability amplitude changed
from $\Delta V=0.2$~mag to $\Delta V=0.8$~mag. The shape of the phased
light curve also changed (see Fig.~7). Star's rotational period is equal to
5.029~d and it is constant throughout all observational seasons what may
suggest the presence of a close companion star. The star is very active
($\log (R_x)=-2.7$) and its X-ray emission is hard (${\it HR1}=0.80$). We
consider this star as a RS CVn type star candidate. Photometric
measurements and a phased light curves for seasons of modest and largest
brightness variations are presented in Fig.~7.

Rotationally variable stars with large photometric variability are
sometimes misidentified with pulsating stars or eclipsing variables. There
are two stars in ACVS with maximum amplitudes of luminosity changes above
$\Delta V=0.5$~mag and properties very similar to the ASAS
063656-0521.0. They are: CD-58 693 (ASAS 032738-5809.5), identified as a
Cepheid with the period of $P=4.075$~d, and ASAS 131055-4844.0, classified
as as a probable spotted contact binary with two most significant periods
$P_1=7.06562$~d and $P_2=3.537421$~d (Pilecki and Szczygie³ 2007). They are
likely very active ($\log(R_x)\approx-2.8$) RS~CVn type stars, but their
binarity has to be confirmed.

\subsection{Eclipsing Binaries, Ellipsoidal Variable Stars and 
Spectroscopic Binaries}
There are 127 detached eclipsing stars in our list (33 new), 98 binaries
with deformed components but unequal minimum depths (19 new) and 124
contact binaries (11 new).

There are some early type eclipsing binaries with X-ray emission due to the
presence of strong stellar winds (\eg the O-type stars NGC3603 A1, MY~Ser),
but most of eclipsing binaries have spectral type G--K, so their X-ray
emission is related to the coronal activity. Changes in the phased light
curve resulting from the star-spots may influence the distinction between a
detached binary with well separated, almost spherical, components and an
eclipsing binary with tidally deformed stars.  The star HD 184415
(193504-1833.9) with a period of 45.6~d is the longest period eclipsing
binary in our catalog, but the eclipses are shallow and probably masked
sometimes by star spots.

The periods of detached eclipsing binaries in our sample are between
0.37069~d (ASAS 210445-3525.9) and 43.5~d (ASAS 065114+0754.1). Among
detached eclipsing binaries there are two very young stars, as suggests
their spectrum and strong lithium line: TYC 8283-264-1 (ASAS 144135-4700.6,
orbital period 2.0171 d) and TYC 7310-503-1 (ASAS 145837-3540.5, orbital
period 2.09789~d). The second of these stars has been the subject of the
recent study by Hebb \etal (2011).

The distinction between contact and near contact binaries based on
photometry is often unreliable, so our classification is preliminary in
this respect and needs to be refined {\it via} modeling and spectroscopic
observations. We are also aware that some stars we consider as contact
binaries or eclipsing binaries with deformed components (especially with
low amplitude brightness variations) may be in fact variable due to the
presence of spots.

Among stars classified as contact binaries there are three outstanding
objects with similar properties. These are red stars ($V-I$ color in the
range 1.6--1.8~mag) with periods close to one day (ASAS 122456-4854.4,
$P=1.1015$~d, $A_I=0.16$~mag, $A_V=0.22$~mag; ASAS 134124-5205.3,
$P=1.35808$~d, $A_I=0.08$~mag, $A_V=0.10$~mag; ASAS 162341-3306.2,
$P=0.99498$~d, $A_I=0.28$~mag), and high X-ray activity with $\log(R_X)$
close to $-3$. It is possible that these stars are single fast rotating
late K or early M-type dwarfs with large stable spots.
 
If we ignore these three stars, the periods of the other contact binaries
from our list range from 0.21781 d (ASAS 083127+1952.9, subject to the
recent study by Ruciñski and Pribulla 2008) up to 0.68145~d (DO~Cha or ASAS
090744-8219.4).

As ellipsoidal variable stars we classified stars with regular brightness
changes, which do not fit to the period--color relation (Eggen 1967) for
classical contact binaries. The tidal deformation of a larger component may
result in brightness changes up to 0.2~mag as in the case of AE~Aqr (ASAS
204009-0052.2) where the presence of the white dwarf is responsible for the
deformation of the main sequence star.

Data found in the literature indicate that 263 non eclipsing stars have
spectroscopic companions (there are also 40 stars with probable
spectroscopic companions). For 201 stars available data indicate constant
radial velocity so we consider these stars as single.

Most of eclipsing stars included in our catalog are in close binary
systems, so the rotation of both components should be synchronized or
almost synchronized with the orbital period. In the subsequent study of the
relation between activity and rotation we will include the eclipsing
binaries with the X-ray emission of coronal origin.

\subsection{Stars with Miscellaneous Variability Classification}
Five stars in our list are classified as miscellaneous variable stars. TYC
8577-1672-1 (ASAS 85008-5745.8) is a blend of a G-type star with an M5III
star. Periodicity of 52.5~d is pronounced only in the {\it I}-band where
M-type star should dominate. KP~Aqr (ASAS 220945-1353.7, spectral type
M2III) with a period of about 22~d may be an example of OSARG type star
(Soszyñski \etal 2011) \ie the red giant with pulsations supported by
convection. The T~Tau type star TYC 6793-819-1 (161410-2305.6) presents a
luminosity changes with variable amplitude and period of 35.8~d. Its
projected rotation velocity (27.5~km/s, T06), implies $R\sin i$ of 19~\RS,
too large for a K2IV star.

Two other stars are cataclysmic variables presenting at the time of ASAS
observations large brightness dispersion and the seasonal modulation of the
mean brightness with a period of 10~d (V341~Ara) and 17~d (IX~Vel).

Variability of V341~Ara has already been reported many times. The star was
initially classified as a type II Cepheid with a period of
11.8~d. Berdnikov and Szabados (1998) found the period of 14.11~d, and {\it
V} amplitude of about 0.5~mag, whereas the period found in the Hipparcos
photometry (Perryman \etal 1997) was $P=10.92$~d. V341~Ara has a blue color
($V-I=0.18$~mag) and O-type spectral classification, so it is probably a
nova like variable star. During ASAS observations (seasons 2000--2008)
large scatter is present with the amplitude of about 0.5~mag in the {\it
I}- and {\it V}-bands and we obtain clear periodicity only for the last
season of the {\it V} data.  This periodic signal is strong, but the large
residua remain (Fig.~8).
\begin{figure}[htb]
\hglue-7mm{\includegraphics[width=13.5cm, bb=23 440 575 700]{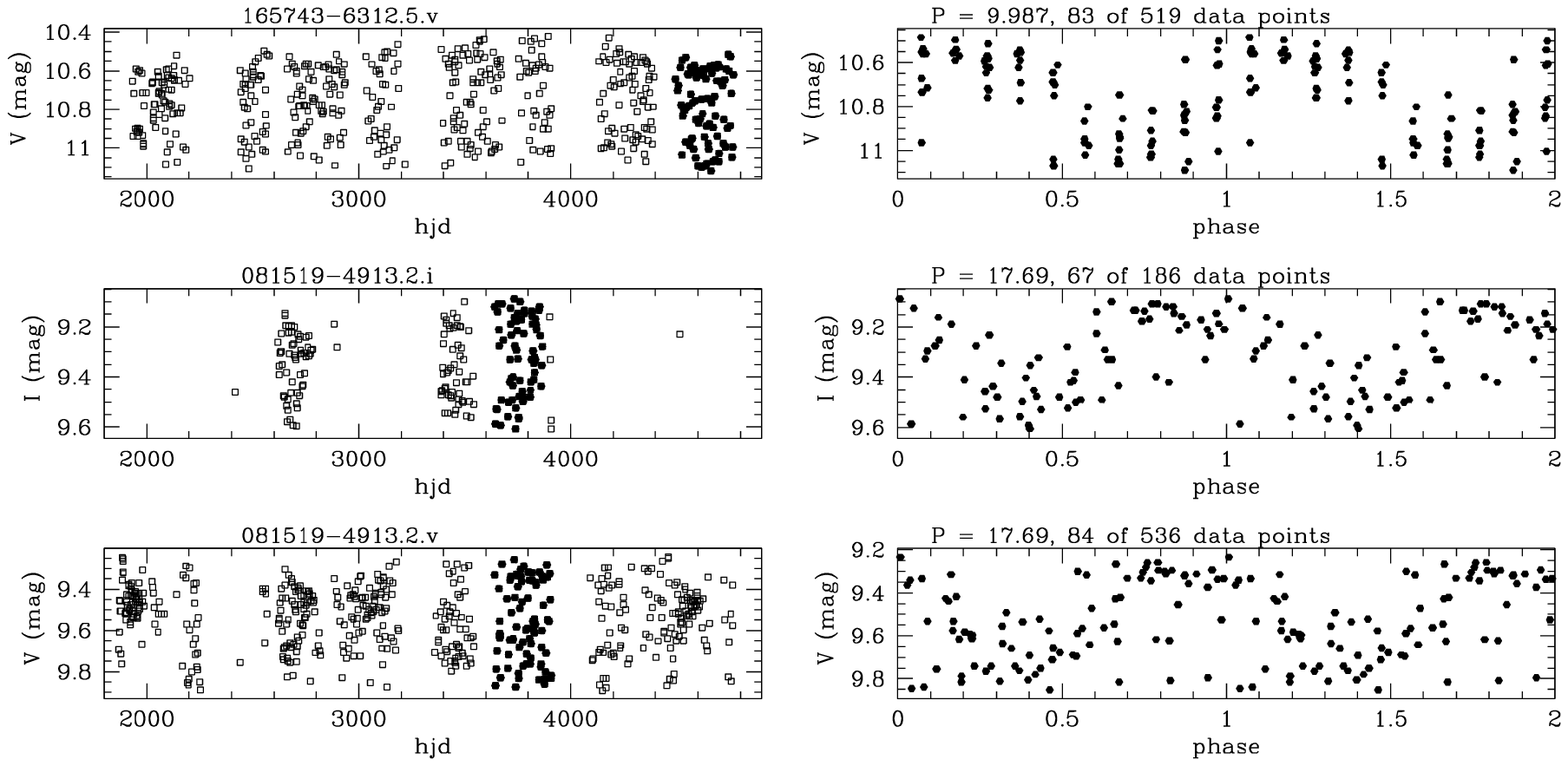}}
\FigCap{Photometric data for two cataclysmic binaries V341~Ara and 
IX~Vel. Photometric data points as a function of ${\rm hjd=HJD-2450000}$
are presented on the {\it left side}. Phased light curve of data showing
periodic variability is presented on the {\it right side}. {\it V}-band
data for V341~Ara is presented in the {\it upper panels}, {\it I}-band and
{\it V}-band data for IX~Vel are presented in the {\it middle panels} and
{\it lowermost panels}, respectively.}
\end{figure}

IX~Vel (ASAS 081519-4913.2) is a well studied nova like cataclysmic
variable star with the orbital period of $P_{\rm orb}=0.193929$~d. White
dwarf has a mass about 0.8~\MS\ and effective temperature of about
60\,000~K. Secondary component fills its Roche lobe and its mass is about
0.5~\MS\ (Linnell \etal 2007). The star has properties similar to
V341~Ara. Both have blue color ($V-I\approx0.2$~mag), similar level of the
X-ray activity ($\log(R_{x})=-3.15$, although the X-ray spectrum of
V341~Ara is harder) and the large scatter in the observed brightness (both
in the {\it I}- and {\it V}-bands). There are observing seasons (see
Fig.~8) when the large amplitude scatter in the photometric data
(flickering?) is modulated with a period of 10~d (V341~Ara) and 17.7~d
(IX~Vel).

\subsection{Projected Rotational Velocities and Minimal Radii of Stars}
We consider the projected rotational velocity as a useful information
enabling a distinction between the fast and slow rotating stars in cases
when photometric data do not eliminate the aliased periods well separated
in the period domain. There are 850 stars in our list with the projected
rotational velocity found in the literature.

\begin{figure}[htb]
\centerline{\includegraphics[width=13cm, bb=23 425 575 700]{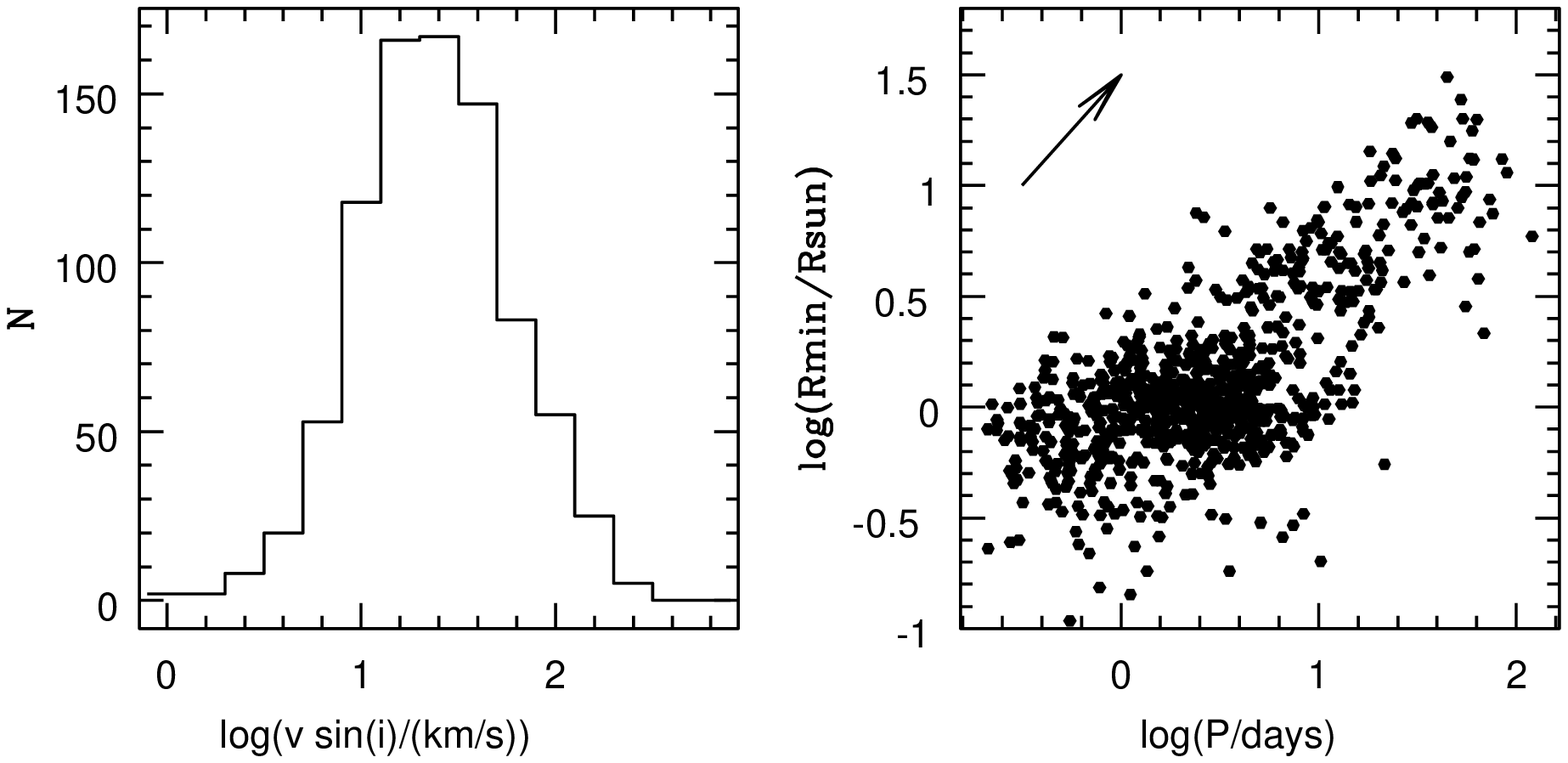}}
\FigCap{{\it Left panel}: histogram presenting number of stars in 
logarithmic bins of projected rotational velocity (the bin size is 0.2).
{\it Right panel}: inferred minimal radii of stars with the known values of
projected rotational velocity plotted as a function of variability
period.}
\end{figure}

Histogram of $\log(\vv\sin i)$ is shown in Fig.~9 (left panel). Most stars
have projected equatorial velocities in the range 10--100~km/s. The largest
projected rotation velocities (over 200~km/s) have been measured for three,
most probably very young, G-type stars. G9 spectral type star V370~Vel (ASAS
084214-5256.1) has the rotation period of $P=0.223$~d (Patten and Simon
1996) and $\vv\sin i=235$~km/s (Marsden \etal 2009). HD 175897 (ASAS
190109-5853.5) has the spectral type G5V, pronounced lithium line and
$\vv\sin i=230$~km/s (T06). For this star we have found rotation period
$P=0.4587$~d indicating that this fast rotating star is still in the
pre-main sequence phase ($R\sin i=2.1~\RS$). BD-14 5534 (ASAS
194536-1427.9) has similar properties: spectral type G6V, $\vv\sin
i=206$~km/s (T06), and the rotation period equal to 0.5077~d.

The value of $R\sin i$ as a function of rotational period for stars with
measured projected rotational velocity is presented in the right side of
Fig.~9.  There are some cases where the value of the projected rotational
velocity seems to contradict the measured rotational period of the star.
For example the star CD-45 14393 (ASAS 213816-4511.5) has the spectral type
K3Ve, so its radius should be about 0.7~\RS. The projected rotational
velocity is 10~km/s (with a formal uncertainty of 2~km/s -- T06), whereas
the rotational period obtained from photometry seems to be 0.553~d (1.242~d
alias is less probable). This indicates $R\sin i=0.11$~\RS\ (0.25~\RS\ for
$P=1.24$~d) which implies a low inclination of the rotation axis to the
line of sight (about 10 degrees for $P=0.553$~d and about 20 degrees for
$P=1.242$~d). Amplitude of the photometric variability is 0.13 mag in {\it
V}-band which contradicts the low inclination because it implicates very
large brightness variations in the equatorial band.

\subsection{Stellar Parallaxes and Kinematic Data}
Proper motions are listed for 1632 stars from our sample. The star with the
largest proper motion, equal to 850~mas/yr, is HD 196998 (ASAS
204142-2219.1). Stars with still larger proper motion were missing in
different epochs of ROSAT and ASAS data.

Radial velocity is given for 1092 stars. Histogram representing the number
of stars in 4~km/s bins is shown in Fig.~10. Although there are several
strong peaks in the histogram, only peaks around 0~km/s can be easily
identified with the Lupus--Centaurus--Scorpius association. There are four
stars with the radial velocity smaller than $-100$~km/s, and three with the
radial velocity larger than 100~km/s. The most outstanding example is
V474~Car (ASAS 090022-6300.1) with radial velocity almost 250~km/s, what
makes it a strong candidate for Population~II star.
\begin{figure}[htb]
\centerline{\includegraphics[width=7cm]{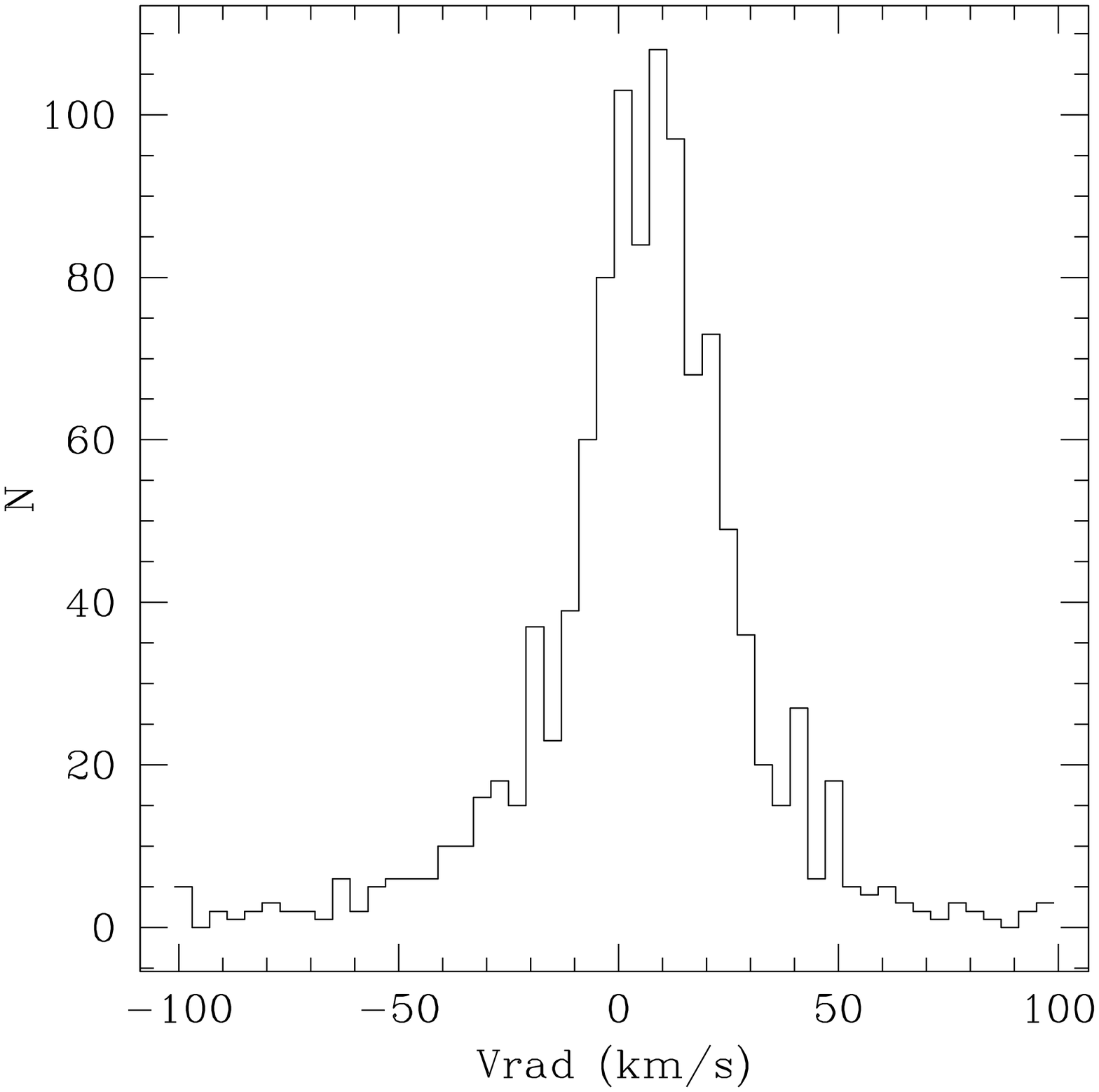}}
\FigCap{Histogram of radial velocities for 1092 variable stars based 
on the data found in literature. The bin size is 4~km/s. Note that part of
data is based on a single epoch measurement.}
\end{figure}
\begin{figure}[htb]
\centerline{\includegraphics[width=13cm, bb=23 425 575 700]{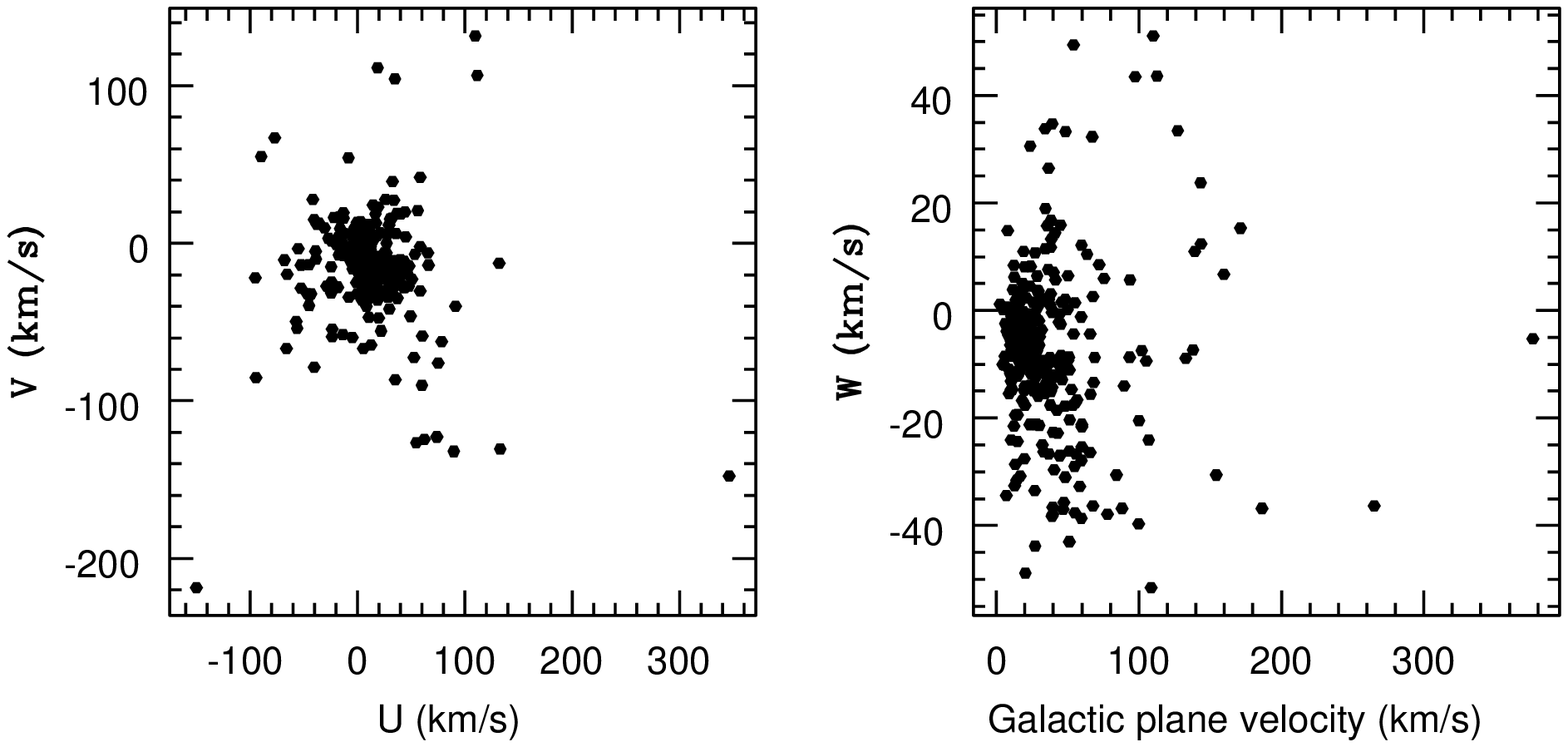}}
\FigCap{{\it U} and {\it V} components of space velocity for stars with given
parallax, proper motion, and radial velocity ({\it left side}). Component of
space velocity perpendicular to the Galactic plane ($W$) \vs velocity
in the Galactic plane calculated as $(U^2+V^2)^{1/2}$ ({\it right side}).}
\end{figure}

There are 300 stars with full kinematic data \ie parallax, proper motion,
and radial velocity. We plot their {\it U,V,W} velocities in Fig.~11.
There is a large concentration of stars with small velocity relative to the
local interstellar medium (young Galactic disk stars), and less numerous
stars with the high velocities especially in the Galactic plane. Velocities
perpendicular to the Galactic plane are in the range $(-52,+52)$~km/s, but
velocities parallel to the Galactic plane are quite large for some stars.
Two stars have velocities parallel to Galactic plane significantly larger
than others: DR~Oct (ASAS 100739-8504.6) has $(U,V,W)=(346,-148,-5)$~km/s,
and V474~Car (ASAS 090022-6300.1) has $(U,V,W)=(-150,-219,-36)$~km/s.
DR~Oct is a SB1 star with a low metal content ($P_{\rm orb}=5.5739$~d,
${\rm [Fe/H]=-2.4}$, Ardeberg and Lindgren 1991), and its photometric
period is very close to orbital one ($P=5.574$~d, Cutispoto 1998) and
variability typical for the presence of spots. The Population~II RS~CVn
type star V474~Car was a subject of a recent study by Bubar \etal (2011).
It is also an SB1 star with the orbital period equal to 10.19~d and the
ASAS photometric period of 10.26~d.

\subsection{X-ray Activity and Objects with the Largest X-ray to Bolometric 
Luminosity Ratio} 
The overall period--activity relation for all periodic variable stars in
our sample is presented in Fig.~12. The histogram of $\log (R_x)$ is
presented in the left panel of Fig.~13.

\begin{figure}[htb]
\vspace*{-15pt}
\centerline{\includegraphics[width=7cm]{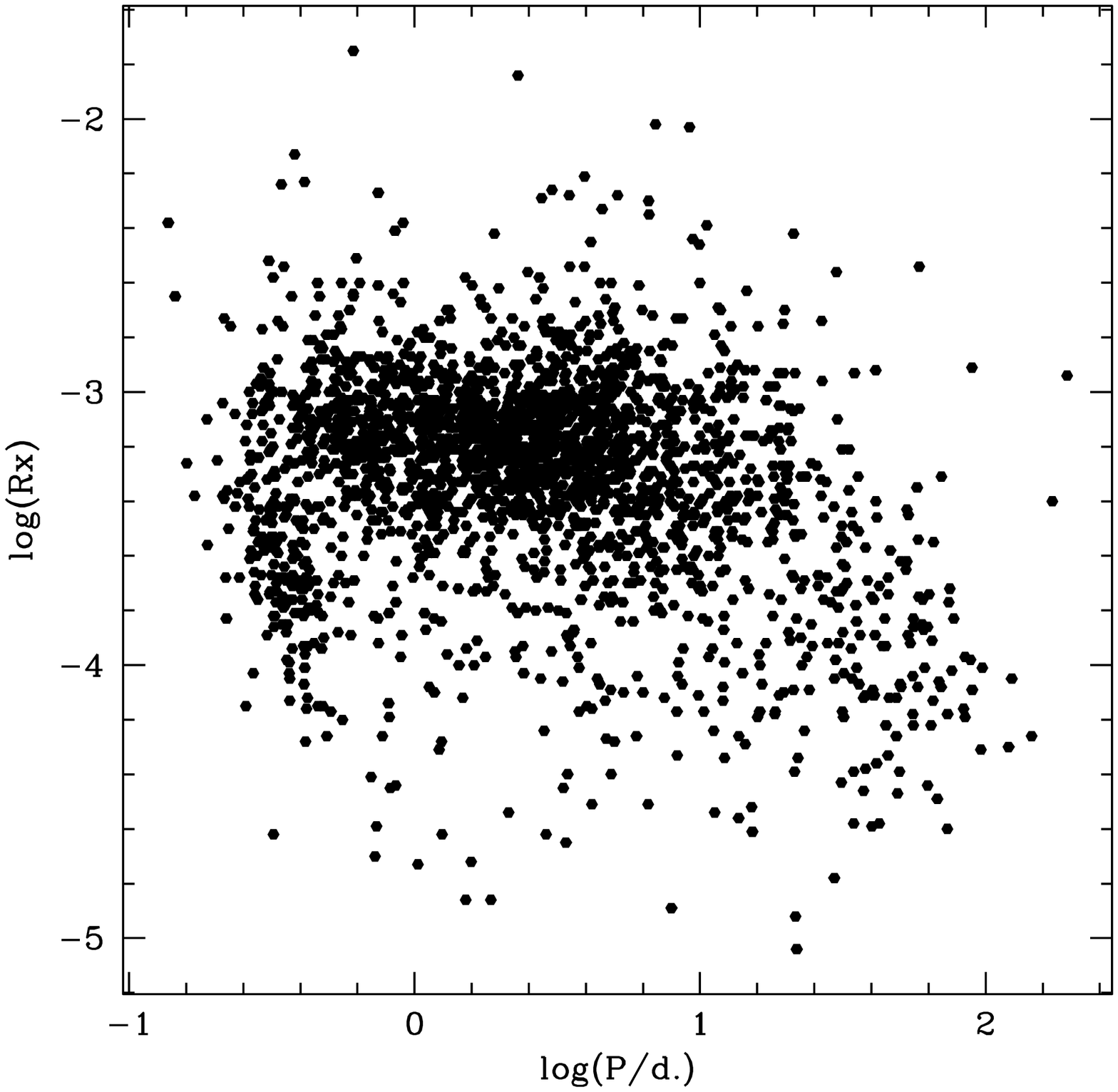}}
\FigCap{X-ray to bolometric flux ratio given as a function of
variability period for stars listed in the catalog.}
\end{figure}
\begin{figure}[htb]
\centerline{\includegraphics[width=12.5cm, bb=23 420 570 700]{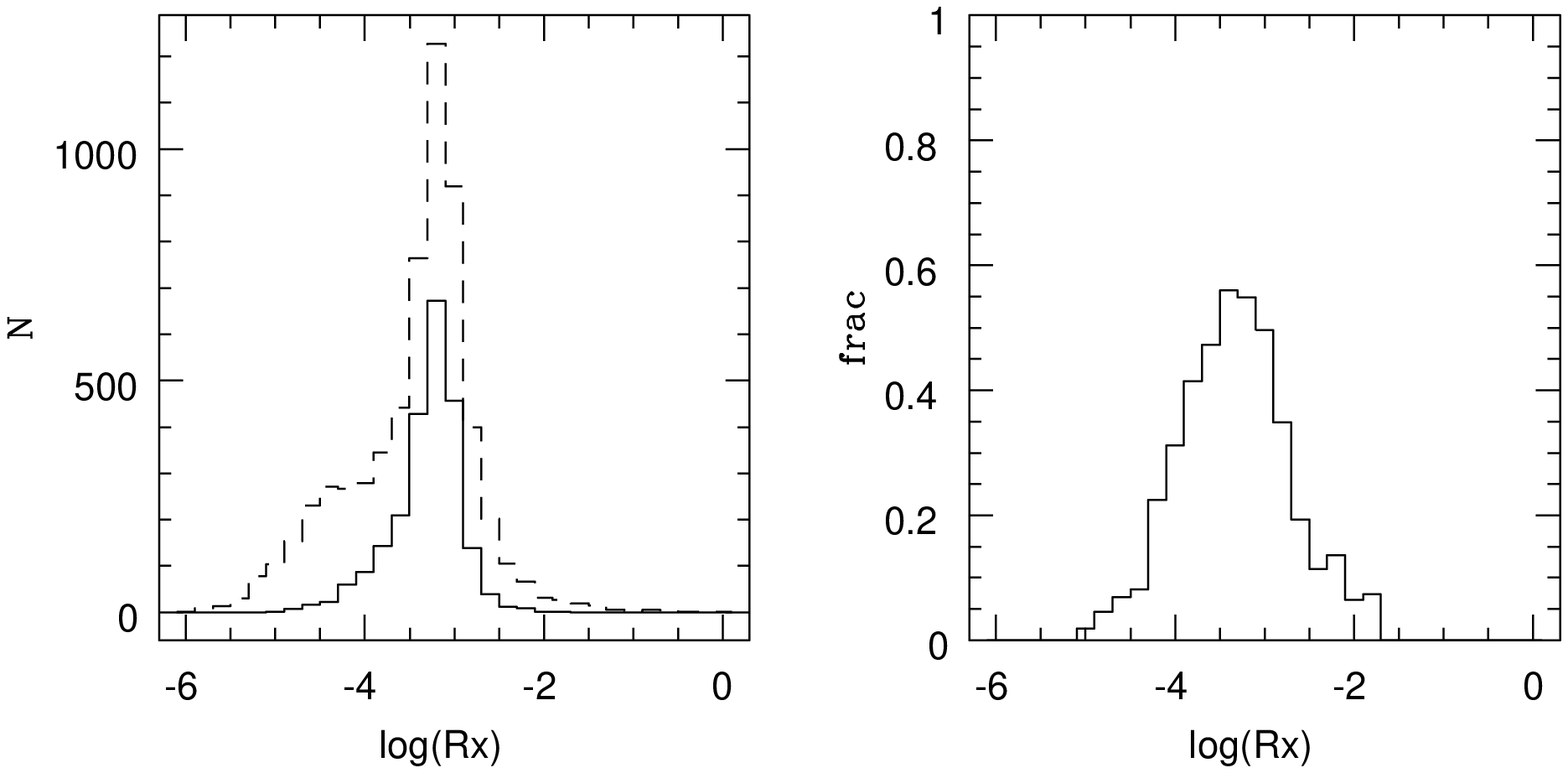}}
\FigCap{{\it Left panel}: number of stars in $\log (R_x)$ bins equal to
0.25. A solid line indicates 2302 periodic variable stars and a dotted line
indicates 6028 stars we investigated for periodic luminosity changes. {\it
Right panel}: fraction of variable stars relative to all investigated stars
as a function of X-ray activity expressed as $\log (R_x)$.}
\end{figure}
The saturation of coronal X-ray emission at the level of $\log (R_x)=-3.0$
is very well established (\eg G\"udel 2004) and it is clearly visible in our
data. There are 377 variable objects with $\log (R_x)>-3.0$, only 25 with
$\log (R_x)>-2.5$ and 2 with $\log (R_x)>-2.0$. The X-ray emission of two
most active stars is not of coronal origin but it is related to the
presence of white dwarfs (WD). BV~Cen is a cataclysmic binary with an
accreting WD and TYC~8769-1214-1 has a close (angular distance
0\zdot\arcs2) young hot WD companion.
 
Among ten variable stars with the highest $R_x$ four are related to white
dwarfs, (BV~Cen, TYC~8769-1214-1, AE~Aqr, FF~Aqr), three are rotationally
variable red dwarfs characterized by $V-I>2$~mag (ASAS designations:
031523-2821.4, 032439-3904.3, 112548-4410.4), one is an eclipsing binary
(RS~CVn type star ASAS 080721-1103.6), one is a pre-main sequence star
BO~Mic (``Speedy Microscopium''), and one (ASAS 121616-5055.3) is a blend
of two stars of similar brightness.

We are aware that formula derived for X-ray coronal emission may be not
adequate for calculation of the X-ray flux and $R_x$ for other type sources
like accreting or thermally emitting white dwarfs or wind colliding
binaries. Very high X-ray to bolometric luminosity ratio for some M-type
dwarfs may be related to flares.

The detailed study of activity--rotation relation for various classes of
coronally active stars will be given in the forthcoming papers.

\subsection{Fraction of Variable Stars}
We were surprised by high fraction of variable stars in our study: 2302 out
of 6028 investigated.

Rotational variability of stars is tied to their coronal activity. Most of
stars we analyzed for variability have $\log(R_x)$ in the range $(-4,-3)$
and almost 50\% of stars in this activity range are variable (see Fig.~13).
Activity of $\log (R_x) > -2$ is too high to be of coronal origin so we do
not expect variability due to spots. In addition, objects with the highest
X-ray to bolometric luminosity ratio are relatively faint, so their
photometry is less precise. Lower activity on the other hand is related to
the lower magnetic field filling factor, smaller star-spots and weaker
variability.

The fraction of the detected variable stars depends to some degree on
stellar declination due to the number of available observations. The mean
number of observations in the {\it V}-band decreases gradually between the
declination $-75\arcd$ and $25\arcd$ from 800 to about 250 measurements per
star. The mean number of observations per star in the {\it I}-band is
roughly constant at the level of 250 measurements per star between
$-60\arcd$ and $20\arcd$ but it drops significantly for declinations
greater than $20\arcd$. The fraction of stars we found as variable
drops from about 45\% for declinations smaller than $-40\arcd$ to 25\% for
declinations larger than $20\arcd$.

\subsection{Comparison with Earlier Studies}
Our variable star list includes many objects known to be variable and
included in many previous studies and catalogs. For example we have 58
stars common with Hipparcos survey (KE2002), 76 with SuperWASP variable
star study (N07), 731 with ASAS Catalog of Variable Stars
(ACVS). Altogether we list 1001 stars with already known photometric or
orbital periods, although in many cases our periods are different (mostly
due to aliases or a choice of different harmonics).

A brief comparison of our results with N07 was given in the Section~2.1, so
here we shortly compare our results with KS2007, KE2002 and ACVS.
 
{\it Comparison to Kiraga and Stêpieñ (2007)}\\ Our detection threshold in
the present study was limited to the cases when AoV statistics for some
part of data was larger than 10, so it was more restrictive than in our
previous study of M dwarfs where it was set to 8.  We analyzed 26 of 31
stars listed as variable in KS2007 and found 18 of them to be variable. For
12 stars, periods listed in the present catalog and periods of KS2007 are
similar.

In six cases we list different periods (aliases of the periods given by
KS2007), but in three cases we decided to include the periods given by
KS2007 to the list of possible aliases.

{\it Comparison with Koen and Eyer (2002)}\\
We calculated AoV statistics for 110 objects present in the KE2002 catalog
and 58 are included in our list. We obtained similar periods in 29 cases
(50 \%). In 10 cases different periods cannot be attributed to aliases or
different harmonics.

{\it Comparison with ACVS (ASAS Catalogue of Variable Stars)}\\
We computed AoV statistics for 809 stars coincident with ACVS entries and
731 were included into our final list. Periods of 518 stars are similar,
for 92 stars differ by a factor of two, and for 105 they are aliases. For
16 stars the differences are of different origin. The main difference
between our work and ACVS is in classification of variable stars.
  
\Section{Summary and Conclusions}
We analyzed bright ($8<I<12.5$~mag or $8<V<12.5$~mag) sources from the ASAS
-- South photometric catalog coincident with the sources from the ROSAT All
Sky Survey (Voges \etal 1999) and found 2302 of 6028 stars to be variable.
The majority of stars have $V-I$ colors and spectral classification typical
for coronally active stars possessing convective envelopes, but several
cataclysmic variable stars and early type binaries are also present.

We have 731 stars common with ASAS Catalog of Variable Stars (ACVS), 76
stars with Norton \etal (2007) and 58 stars common with Koen and Eyer
(2002). Altogether 993 photometric periods and 8 spectroscopic ones were
found in the literature. Aliasing appears to be the main uncertainty of
the obtained periods.

Because of the small angular resolution of ASAS cameras (one pixel equals
to 14\arcs) we inspected DSS images for possible blended stars. We found
139 companions of comparable brightness and 336 fainter ones. Additional
information about 150 close neighbors was found in the literature and
catalogs {\it via} SIMBAD database. There are 1679 variable stars on our
list without significant blending.

Many stars are suspected to be of RS~CVn type. There are 127 stars
classified as detached eclipsing binaries and 220 stars classified as close
eclipsing binaries with deformed components. Data found in the literature
indicate that 263 stars not classified as eclipsing binaries have close
spectroscopic companions (there are also 40 stars with probable
spectroscopic companions). For 201 stars available data indicate 
constant radial velocity so these stars are most probably single.

For many stars we did not find additional information in the literature and
our classification based on photometric data should be considered as
preliminary only. In many cases it would be necessary to obtain
spectroscopic data to confirm or reject the binary nature of stars listed
in the present catalog. In the case of stable variability period, amplitude
and phase, we considered a star to be binary. We assume that changes of
amplitude indicate the presence of star-spots.

We presented several stars with outstanding properties. Some of them have
the X-ray emission of non coronal origin. Most new variable stars have
small amplitudes, but there are also examples of stars with very large
brightness variations. The star ASAS 063656-0521.0 has a particularly
large amplitude due to presence of star-spots (up to $\Delta V=0.8$~mag,
Section~4.4).
 
The present catalog contains data useful for study of activity--rotation
relation for several classes of active stars (young stars, main sequence
stars, evolved stars, single stars, close binaries), which will be
presented in the forthcoming papers.

\Acknow{
We are particularly grateful to Prof. Grzegorz Pojmañ\-ski for help in using
ASAS data, help in improving manuscript, and many useful comments, and
Prof. Kazimierz Stêpieñ for discussions related to a subject of this work.
This research has made use of the SIMBAD database, operated at CDS,
Strasbourg, France}

\end{document}